\documentclass[aip,jcp,preprint,groupedaddress]{revtex4-2}%
\usepackage{graphicx}
\usepackage{tabularx}
\usepackage{dcolumn}
\usepackage{longtable}
\usepackage{tensor}
\usepackage{color}
\usepackage{placeins}
\usepackage{bm}
\usepackage{amsmath}
\usepackage{amsfonts}
\usepackage{amssymb}%
\providecommand{\U}[1]{\protect\rule{.1in}{.1in}}
\providecommand{\U}[1]{\protect\rule{.1in}{.1in}}
\providecommand{\U}[1]{\protect\rule{.1in}{.1in}}
\begin{document}
\title{How important are the residual nonadiabatic couplings for an accurate
simulation of nonadiabatic quantum dynamics in a quasidiabatic representation?}
\author{Seonghoon Choi}
\email{seonghoon.choi@epfl.ch}
\author{Ji\v{r}\'{\i} Van\'{\i}\v{c}ek}
\email{jiri.vanicek@epfl.ch}
\affiliation{Laboratory of Theoretical Physical Chemistry, Institut des Sciences et
Ing\'enierie Chimiques, Ecole Polytechnique F\'ed\'erale de Lausanne (EPFL),
CH-1015, Lausanne, Switzerland}
\date{\today}

\begin{abstract}
Diabatization of the molecular Hamiltonian is a standard
approach to removing the singularities of nonadiabatic couplings at conical
intersections of adiabatic potential energy surfaces. In general,
it is impossible to eliminate the nonadiabatic couplings entirely---the
resulting \textquotedblleft quasidiabatic\textquotedblright\ states are still
coupled by smaller but nonvanishing residual nonadiabatic couplings, which are
typically neglected. Here, we propose a general method for assessing the
validity of this potentially drastic approximation by comparing quantum
dynamics simulated either with or without the residual couplings. To make the
numerical errors negligible to the errors due to neglecting the residual
couplings, we use the highly accurate and general eighth-order composition of
the implicit midpoint method. The usefulness of the proposed
method is demonstrated on nonadiabatic simulations in the cubic Jahn--Teller
model of nitrogen trioxide and in the induced Renner--Teller model of hydrogen
cyanide. We find that, depending on the system, initial state, and employed
quasidiabatization scheme, neglecting the residual couplings can result in
wrong dynamics. In contrast, simulations with the exact
quasidiabatic Hamiltonian, which contains the residual couplings, always yield
accurate results.

\end{abstract}
\maketitle

\graphicspath{{./figures/}{C:/Users/Jiri/Dropbox/Papers/Chemistry_papers/2020/Residual_coupling/figures/}{"d:/Group Vanicek/Desktop/Choi/Residual_coupling/figures/"}}

\section{Introduction}

The celebrated Born--Oppenheimer approximation,\cite{Born_Oppenheimer:1927}
which treats the electronic and nuclear motions in molecules separately, is no
longer valid for describing processes involving two or more strongly
vibronically coupled electronic states. A common approach that goes beyond
this approximation\cite{Domcke_Yarkony:2012, book_Nakamura:2012,
book_Takatsuka:2015, Bircher_Rothlisberger:2017, Shin_Metiu:1995,
Albert_Engel:2016, Abedi_Gross:2010, Cederbaum:2008} consists in solving the
time-dependent Schr\"{o}dinger equation with a truncated molecular Hamiltonian
that includes only a few, most significantly
coupled\cite{Zimmermann_Vanicek:2010,Zimmermann_Vanicek:2012}
Born--Oppenheimer electronic states.\cite{Worth_Cederbaum:2004,
book_Baer:2006, Cederbaum:2004} The \textquotedblleft
adiabatic\textquotedblright\ states, obtained directly from the electronic
structure calculations, are, however, not adequate for representing the
molecular Hamiltonian in the region of strong nonadiabatic couplings; in
particular, the couplings between the states diverge at conical
intersections,\cite{Teller:1937, Herzberg_LonguetHiggins:1963, Zimmerman:1966,
Forster:1970, Domcke_Yarkony:2012, Cederbaum:2004, Yarkony:2004} where
potential energy surfaces of two or more adiabatic states intersect.

\emph{Quasidiabatization}, i.e., a coordinate-dependent unitary
transformation\cite{Koppel:2004, Pacher_Koppel:1989, Pacher_Koppel:1993} of
the molecular Hamiltonian that reduces the magnitude of the nonadiabatic
vector couplings, rectifies this singularity. The transformation matrix can
be obtained by various quasidiabatization schemes, of which a few
representative examples include methods based on the integration of the
nonadiabatic couplings\cite{Baer:1975, Das_Baer:2011, Richings_Worth:2015,
Sadygov_Yarkony:1998, Esry_Sadeghour:2003, Evenhuis_Collins:2004,
Guan_Yarkony:2019, Malbon_Yarkony:2015, Zhu_Yarkony:2010, Zhu_Yarkony:2012,
Zhu_Yarkony:2016} or on different molecular properties\cite{Mulliken:1952,
Hush:1967, Cave_Newton:1997, Werner_Meyer:1981, Yarkony:1998,
Hirsch_Petrongolo:1990, Peric_Buenker:1990} and the
block-diagonalization\cite{Pacher_Koppel:1988, Pacher_Cederbaum:1991,
Neville_Schuurman:2020, Domcke_Woywod:1993} or regularized
diabatization\cite{Thiel_Koppel:1999,Koppel_Mahapatra:2001,
Koppel_Schubert:2006} schemes.

In systems with more than one nuclear degree of freedom, the
\emph{strict diabatization}, which eliminates the nonadiabatic couplings
completely, is only possible if infinitely many electronic states are
considered.\cite{Mead_Truhlar:1982, Pacher_Koppel:1989} The best one can do
for a general subsystem with a finite number of electronic states is the
above-mentioned \emph{quasidiabatization}, in which the unitary transformation
reduces the magnitude of the couplings but does not remove them entirely.
However, it is a common practice to neglect these nonvanishing
\textquotedblleft residual\textquotedblright\ couplings present in the
\emph{exact quasidiabatic Hamiltonian} and thus obtain an \emph{approximate
quasidiabatic Hamiltonian}, whose additional benefit is a simpler, separable
form convenient for quantum simulations.

Here, we propose a general method that quantifies the importance of the
residual couplings by comparing nonadiabatic simulations performed either with
the exact quasidiabatic Hamiltonian---obtained through an exact unitary
transformation of the adiabatic Hamiltonian---or with the approximate
quasidiabatic Hamiltonian, which neglects the residual couplings. By
definition and regardless of the magnitude of the residual couplings, the
results obtained with the exact quasidiabatic Hamiltonian can serve as the
exact benchmark, as long as the numerical errors are
negligible.\cite{Choi_Vanicek:2020} Therefore, for a valid comparison, one
needs a time propagation scheme that can treat even the nonseparable exact
quasidiabatic Hamiltonian and that ensures that the numerical errors are
negligible to the errors due to neglecting the residual couplings.
Among various integrators\cite{Tal-Ezer_Kosloff:1984,
Lanczos:1950, Tal-Ezer:1989, Park_Light:1986, Choi_Vanicek:2019,
Leforestier_Kosloff:1991, book_Leimkuhler_Reich:2004, book_Hairer_Wanner:2006}
that satisfy both requirements, we chose the optimal
eighth-order\cite{Kahan_Li:1997} composition\cite{Suzuki:1990, Yoshida:1990,
book_Hairer_Wanner:2006, book_Lubich:2008} of the implicit midpoint
method\cite{book_Hairer_Wanner:2006, book_Leimkuhler_Reich:2004,
McCullough_Wyatt:1971} because it also preserves
exactly\cite{Choi_Vanicek:2019} various geometric properties of the exact
solution.\cite{book_Hairer_Wanner:2006, book_Leimkuhler_Reich:2004}

After presenting the general method in Sec.~\ref{sec:theory}, in
Sec.~\ref{sec:numerical_examples} we provide realistic numerical examples, in
which we employ the method to quantify the importance of the residual
couplings in nonadiabatic simulations of nitrogen trioxide (NO$_{3}%
$)\cite{book_Bersuker_Polinger:2012, Viel_Eisfeld:2004, Bersuker:2001,
Hauser_Ernst:2010} and hydrogen cyanide (HCN).\cite{Pacher_Koppel:1988,
Koppel_Von_Niessen:1979, Cederbaum_Domcke:1981, Koppel_Cederbaum:1981} Whereas
the NO$_{3}$ model was quasidiabatized with the regularized diabatization
scheme,\cite{Thiel_Koppel:1999, Koppel_Mahapatra:2001, Koppel_Schubert:2006}
the block-diagonalization scheme\cite{Pacher_Koppel:1988,
Pacher_Cederbaum:1991, Neville_Schuurman:2020, Domcke_Woywod:1993} was
employed in the HCN model. To find out how the errors due to ignoring the
residual couplings depend on the sophistication of the quasidiabatization and
on the initial state, in Sec.~\ref{sec:no3_disp} we compare the
first- and second-order regularized diabatization
schemes\cite{Thiel_Koppel:1999, Koppel_Mahapatra:2001, Koppel_Schubert:2006}
on the model of a displaced excitation of NO$_{3}$.

\section{\label{sec:theory}Theory}

We begin by introducing the standard molecular Hamiltonian $\mathcal{H}%
=\mathcal{T}_{\mathrm{N}}+\mathcal{T}_{\mathrm{e}}+\mathcal{V}$, where
$\mathcal{T}_{\mathrm{N}}$ and $\mathcal{T}_{\mathrm{e}}$ are the kinetic
energy operators of the nuclei and electrons, and $\mathcal{V}$ is the
molecular potential energy operator. One may express the molecular Hamiltonian
equivalently as $\mathcal{H}=\mathcal{T}_{\mathrm{N}}+\mathcal{H}_{\mathrm{e}%
}$ by defining the electronic Hamiltonian $\mathcal{H}_{\mathrm{e}%
}:=\mathcal{T}_{\mathrm{e}}+\mathcal{V}$, an operator acting on the electronic
degrees of freedom and depending parametrically on the nuclear coordinates,
described by a $D$-dimensional vector $Q$. For each fixed nuclear geometry,
the time-independent Schr\"{o}dinger equation
\begin{equation}
\mathcal{H}_{\mathrm{e}}(Q)|n(Q)\rangle=V_{n}(Q)|n(Q)\rangle
\label{eq:H_el_tise}%
\end{equation}
for $\mathcal{H}_{\mathrm{e}}(Q)$ can be solved to obtain the $n$th adiabatic
electronic state $|n(Q)\rangle$ and potential energy surface $V_{n}(Q)$ for $n\in%
%TCIMACRO{\U{2115} }%
%BeginExpansion
\mathbb{N}
%EndExpansion
$.

The adiabatic electronic eigenstates $| n (Q) \rangle$, which depend on the nuclear coordinates $Q$, form a complete orthonormal set and can be employed to
expand the exact solution of the time-dependent molecular Schr\"{o}dinger
equation
\begin{equation}
i \hbar\frac{\partial}{\partial t} | \Psi(Q,t) \rangle= \mathcal{H} |
\Psi(Q,t) \rangle\label{eq:tdse}%
\end{equation}
with Hamiltonian $\mathcal{H}$ as an infinite series
\begin{equation}
|\Psi(Q,t)\rangle_{\mathrm{exact}} = \sum_{n = 1}^{\infty} \psi_{n}%
^{\mathrm{ad}}(Q,t) | n (Q) \rangle. \label{eq:born_huang}%
\end{equation}
Note that Eqs.~(\ref{eq:tdse}) and (\ref{eq:born_huang}) combine the
coordinate representation for the nuclei with the representation-independent
Dirac notation for the electronic states; $\psi_{n}^{\mathrm{ad}}(Q,t)$ is the
time-dependent nuclear wavefunction (a wavepacket) on the $n$th adiabatic
electronic surface. The Born--Huang expansion\cite{book_Born_Huang:1954} of
Eq.~(\ref{eq:born_huang}) is exact when an infinite number of electronic
states are included, but in practice, $|\Psi(Q,t)\rangle_{\mathrm{exact}}$ is
approximated by truncating the sum in Eq.~(\ref{eq:born_huang}) and including
only the most important $S$ electronic states:\cite{book_Baer:2006,
Zimmermann_Vanicek:2010, Zimmermann_Vanicek:2012}
\begin{equation}
|\Psi(Q,t)\rangle_{\mathrm{exact}} \approx| \Psi(Q,t) \rangle_{\mathrm{trunc}}
:= \sum_{n = 1}^{S} \psi_{n}^{\mathrm{ad}}(Q,t) | n (Q) \rangle;
\label{eq:born_huang_truncated}%
\end{equation}
for brevity, we shall omit the subscript ``trunc'' in $| \Psi(Q,t)
\rangle_{\mathrm{trunc}}$ from now on.

Substituting ansatz~(\ref{eq:born_huang_truncated}) into the time-dependent
Schr\"{o}dinger equation~(\ref{eq:tdse}) and projecting onto states $\langle
m(Q)|$ for $\ m\in\{1,\dots,S\}$ leads to the ordinary differential equation
\begin{equation}
i\hbar\frac{d}{dt}\bm{\psi}^{\mathrm{ad}}(t)=\hat{\mathbf{H}}_{\mathrm{ad}%
}\bm{\psi}^{\mathrm{ad}}(t),
\end{equation}
expressed in a compact, representation-independent matrix notation:
\textbf{bold} font indicates either an $S\times S$ matrix (i.e., an electronic
operator) or an $S$-dimensional vector, and the hat ($\,\hat{}\,$) denotes a
nuclear operator. In particular, $\hat{\mathbf{H}}_{\mathrm{ad}}$ is the
adiabatic Hamiltonian matrix with elements $(\hat{\mathbf{H}}_{\mathrm{ad}%
})_{mn}=\langle m|\mathcal{H}|n\rangle$, and $\bm{\psi}^{\mathrm{ad}}(t)$ is
the molecular wavepacket in the adiabatic representation with components
$\psi_{n}^{\mathrm{ad}}(t)$. Assuming the standard form $\mathcal{T}%
_{\mathrm{N}}=\hat{P}^{2}/2M$ of the nuclear kinetic energy operator, the
\emph{adiabatic Hamiltonian} matrix is given by the
formula\cite{book_Takatsuka:2015, Domcke_Yarkony:2012, Yarkony:1996,
Cederbaum:2004, Pacher_Koppel:1989}
\begin{equation}
\hat{\mathbf{H}}_{\mathrm{ad}}=\frac{1}{2M}[\hat{P}^{2}\mathbf{1}%
-2i\hbar\mathbf{F}_{\mathrm{ad}}(\hat{Q})\cdot\hat{P}-\hbar^{2}\mathbf{G}%
_{\mathrm{ad}}(\hat{Q})]+\mathbf{V}_{\mathrm{ad}}(\hat{Q}),
\label{eq:mol_H_full}%
\end{equation}
which depends on the diagonal adiabatic potential energy matrix $[\mathbf{V}%
_{\mathrm{ad}}(Q)]_{mn}:=V_{n}(Q)\delta_{mn}$, the nonadiabatic vector
couplings $[\mathbf{F}_{\mathrm{ad}}(Q)]_{mn}:=\langle m(Q)|\nabla
n(Q)\rangle$, and the nonadiabatic scalar couplings $[\mathbf{G}_{\mathrm{ad}%
}(Q)]_{mn}:=\langle m(Q)|\nabla^{2}n(Q)\rangle$. The dot ($\cdot$) denotes the
dot product in the $D$-dimensional nuclear vector space, and $P$ is the
canonical momentum conjugate to $Q$. Note that, for simplicity,
the nuclear coordinates have been scaled so that each nuclear degree of
freedom has the same mass $M$ and, therefore, $M$ is a scalar.

In practice, the nonadiabatic scalar couplings $\mathbf{G}%
_{\mathrm{ad}}(Q)$ in Eq.~(\ref{eq:mol_H_full}) are often neglected, but this
approximation can cause significant errors;\cite{Cotton_Miller:2017,
Reimers_Hush:2015} for the adiabatic Hamiltonian to be exact, it must include
both $\mathbf{F}_{\mathrm{ad}}(Q)$ and $\mathbf{G}_{\mathrm{ad}}(Q)$.
 In Eqs.~(\ref{eq:born_huang})--(\ref{eq:mol_H_full}), one can
freely choose overall phases of the adiabatic electronic states because both
$|n(Q)\rangle$ and $e^{iA_{n}(Q)}|n(Q)\rangle$ [where $A_{n}(Q)$ are
coordinate-dependent phases] are orthonormalized solutions of
Eq.~(\ref{eq:H_el_tise}). In Ref.~\onlinecite{Choi_Vanicek:2020}, we show how
the choice of $A_{n}(Q)$ affects the nonadiabatic couplings $\mathbf{F}%
_{\mathrm{ad}}(Q)$ and $\mathbf{G}_{\mathrm{ad}}(Q)$; in contrast,
$\mathbf{V}_{\mathrm{ad}}(Q)$ remains unaffected.

The nonadiabatic vector couplings can be re-expressed using the
Hellmann-Feynman theorem as
\begin{equation}
\lbrack\mathbf{F}_{\mathrm{ad}}(Q)]_{mn}=\frac{\langle m(Q)|\nabla
\mathcal{H}_{\mathrm{e}}(Q)|n(Q)\rangle}{V_{n}(Q)-V_{m}(Q)},\quad m\neq n,
\end{equation}
accentuating the singularity of these couplings at a conical
intersection\cite{book_Nakamura:2012, Koppel:2004}---a nuclear geometry
$Q_{0}$ where $V_{m}(Q_{0})=V_{n}(Q_{0})$ for $\ m\neq n$%
.\cite{Domcke_Yarkony:2012, Cederbaum:2004, Yarkony:2004} Moreover, Meek and Levine\cite{Meek_Levine:2016} pointed out that, unlike the singularity of $[\mathbf{F}_{\mathrm{ad}}(Q)]_{mn}$, the singularity in the diagonal elements $[\mathbf{G}_{\mathrm{ad}}(Q)]_{nn}$ of the nonadiabatic scalar couplings is not even integrable over domains containing a conical intersection. Another complication
associated with conical intersections is the geometric phase effect: the sign
change of the real-valued adiabatic electronic state $|n(Q)\rangle$ when
transported along a loop containing a conical
intersection.\cite{Longuet-Higgins_Sack:1958, Mead_Truhlar:1979, Berry:1984,
Mead:1992, Kendrick:2000, Juanes-Marcos_Althorpe:2005, Schon_Koppel:1995,
Ryabinkin_Izmailov:2017, Schon_Koppel:1995, Yarkony_Guo:2019,
Joubert-Dorial_Izmaylov:2013, Malbon_Yarkony:2016, Xie_Yarkony:2019,
Xie_Guo:2017} Although the geometric phase effect can be
effectively incorporated into nonadiabatic simulations in the adiabatic basis
by appropriately choosing the above-mentioned phases $A_{n}(Q)$ so that the
states are single-valued,\cite{Longuet-Higgins_Sack:1958, Mead_Truhlar:1979,
Berry:1984, Mead:1992, Kendrick:2000, Juanes-Marcos_Althorpe:2005,
Malbon_Yarkony:2016, Xie_Yarkony:2019, Xie_Guo:2017, Ryabinkin_Izmailov:2017,
Joubert-Dorial_Izmaylov:2013, Schon_Koppel:1995, Yarkony_Guo:2019} the
numerically problematic singularity remains.\cite{Choi_Vanicek:2020} Yet, both
complications, namely the singularity of the nonadiabatic couplings and the
geometric phase effect, can be avoided simultaneously by transforming the
adiabatic Hamiltonian to the quasidiabatic basis
\begin{equation}
|n^{\prime}(Q)\rangle=\sum_{m=1}^{S}|m(Q)\rangle\lbrack\mathbf{S}(Q)^{\dagger
}]_{mn} \label{eq:qd_el_states}%
\end{equation}
and thus obtaining the \emph{exact quasidiabatic Hamiltonian}
\begin{align}
&  \hat{\mathbf{H}}_{\text{qd-exact}}:=\mathbf{S}(\hat{Q})\hat{\mathbf{H}%
}_{\mathrm{ad}}\mathbf{S}(\hat{Q})^{\dagger}\nonumber\\
&  =\frac{1}{2M}[\hat{P}^{2}\mathbf{1}-2i\hbar\mathbf{F}_{\mathrm{qd}}(\hat
{Q})\cdot\hat{P}-\hbar^{2}\mathbf{G}_{\mathrm{qd}}(\hat{Q})]+\mathbf{V}%
_{\mathrm{qd}}(\hat{Q}),\label{eq:mol_H_accurate}%
\end{align}
where $[\mathbf{V}_{\mathrm{qd}}(Q)]_{mn}:=\langle m^{\prime}(Q)|\mathcal{H}%
_{\mathrm{e}}(Q)|n^{\prime}(Q)\rangle$ is the nondiagonal quasidiabatic
potential energy matrix, while $[\mathbf{F}_{\mathrm{qd}}(Q)]_{mn}:=\langle
m^{\prime}(Q)|\nabla n^{\prime}(Q)\rangle$ and $[\mathbf{G}_{\mathrm{qd}%
}(Q)]_{mn}:=\langle m^{\prime}(Q)|\nabla^{2}n^{\prime}(Q)\rangle$ are the
residual vector and scalar couplings, respectively. Note that the molecular state $|\Psi(Q,t)\rangle$ from Eq.~(\ref{eq:born_huang_truncated}) is independent of the choice of basis because the transformation~(\ref{eq:qd_el_states}) from the adiabatic to quasidiabatic electronic basis is accompanied by a simultaneous transformation 
\begin{equation}
\psi^{\mathrm{qd}}_{n}(Q,t) = \sum_{m=1}^{S} [\mathbf{S}(Q)]_{nm} \psi^{\mathrm{ad}}_{m}(Q,t) \label{eq:qd_nuc_wavef}
\end{equation}
of nuclear wavefunctions.
The transformation matrix
$\mathbf{S}(Q)$ is obtained by any of the many quasidiabatization
schemes,\cite{Koppel:2004, Pacher_Koppel:1988, Pacher_Koppel:1993,
Pacher_Koppel:1989, Pacher_Cederbaum:1991, Neville_Schuurman:2020,
Domcke_Woywod:1993, Baer:1975, Das_Baer:2011, Richings_Worth:2015,
Sadygov_Yarkony:1998, Esry_Sadeghour:2003, Evenhuis_Collins:2004,
Guan_Yarkony:2019, Malbon_Yarkony:2015, Zhu_Yarkony:2010, Zhu_Yarkony:2012,
Zhu_Yarkony:2016, Mulliken:1952, Hush:1967, Cave_Newton:1997,
Werner_Meyer:1981, Yarkony:1998, Hirsch_Petrongolo:1990, Peric_Buenker:1990,
Thiel_Koppel:1999, Koppel_Mahapatra:2001, Koppel_Schubert:2006} but the
magnitude of the residual nonadiabatic couplings depends on the scheme.
Following Ref.~\onlinecite{Pacher_Koppel:1989}, we measure this magnitude with
the quantity
\begin{equation}
\mathcal{R[}\mathbf{F}_{\mathrm{qd}}(Q)\mathcal{]}:=\int\Vert\mathbf{F}%
_{\mathrm{qd}}(Q)\Vert^{2}dQ,\label{eq:mag_F}%
\end{equation}
where
\begin{align}
\Vert\mathbf{F}_{\mathrm{qd}}(Q)\Vert^{2} &  :=\mathrm{Tr}[\mathbf{F}%
_{\mathrm{qd}}(Q)^{\dagger}\cdot\mathbf{F}_{\mathrm{qd}}(Q)]\nonumber\\
&  =\mathrm{Tr}[\sum_{l=1}^{D}\mathbf{F}_{\mathrm{qd}}(Q)_{l}^{\dagger
}\mathbf{F}_{\mathrm{qd}}(Q)_{l}]\label{eq:frib_norm_f}%
\end{align}
is the square of the Frobenius norm of $\mathbf{F}_{\mathrm{qd}}(Q)$ [note
that the evaluation of Eq.~(\ref{eq:frib_norm_f}) involves both a matrix
product of $S\times S$ matrices and a scalar product of $D$-vectors].
Section~S1 of the supplementary material describes the numerical
evaluation of $\mathcal{R}[\mathbf{F}_{\mathrm{qd}}(Q)]$ in further detail.

It is well-known that, unless $S$ is infinite or $D=1$, in a general system no
diabatization scheme yields the strictly diabatic states [i.e., states in
which the exact Hamiltonian (\ref{eq:mol_H_accurate}) has zero residual
nonadiabatic couplings].\cite{Mead_Truhlar:1982, Pacher_Koppel:1989} The
transformation by a finite $S\times S$ matrix $\mathbf{S}(Q)$ can
only lead to quasidiabatic states, which are coupled both by the off-diagonal
($m\neq n$) elements $[\mathbf{V}_{\mathrm{qd}}(Q)]_{mn}$ of the quasidiabatic
potential energy matrix and by the---perhaps small but nonvanishing---residual
nonadiabatic couplings.\cite{Yarkony:1996a} In practice, however, these
residual couplings are often ignored in Eq.~(\ref{eq:mol_H_accurate}) in order
to obtain the \emph{approximate quasidiabatic Hamiltonian}
\begin{equation}
\hat{\mathbf{H}}_{\text{qd-approx}}:=\frac{\hat{P}^{2}}{2M}\mathbf{1}%
+\mathbf{V}_{\mathrm{qd}}(\hat{Q}). \label{eq:mol_H_approx}%
\end{equation}

Although the magnitude $\mathcal{R}[\mathbf{F}_{\mathrm{qd}}(Q)]$
itself may indicate whether it is admissible to neglect the residual
couplings, a much more rigorous way to quantify the impact of this
approximation on a particular nonadiabatic simulation consists in evaluating
the quantum fidelity\cite{Peres:1984}
\begin{equation}
\mathcal{F}(t):=|\langle\bm{\psi}_{\text{qd-approx}}%
(t)|\bm{\psi}_{\text{qd-exact}}(t)\rangle|^{2}\in\lbrack0,1]
\label{eq:fidelity}%
\end{equation}
and distance
\begin{equation}
\mathcal{D}(t):=\Vert\bm{\psi}_{\text{qd-approx}}%
(t)-\bm{\psi}_{\text{qd-exact}}(t)\Vert\in\lbrack0,2] \label{eq:distance}%
\end{equation}
between the states $\bm{\psi}_{\text{qd-approx}}(t)$ and
$\bm{\psi}_{\text{qd-exact}}(t)$, evolved with the approximate and exact
quasidiabatic Hamiltonians, respectively. [I.e., $\bm{\psi}_{i}(t)=\exp
{(-i\hat{\mathbf{H}}_{i}t/\hbar)}\bm{\psi}(0)$ for $i\in\{\text{qd-approx}%
,\text{qd-exact}\}$.] The more important the residual couplings, the smaller
the quantum fidelity and the larger the distance.

By both propagating and comparing the wavepackets $\bm{\psi}_{\text{qd-approx}%
}(t)$ and $\bm{\psi}_{\text{qd-exact}}(t)$ in the same quasidiabatic
representation, one avoids contaminating the errors due to the neglect of the
residual couplings with the numerical errors due to the transformation between
representations. In fact, as long as it is numerically converged,
$\bm{\psi}_{\text{qd-exact}}(t)$ serves as the exact benchmark regardless of
the size of the residual couplings because the exact quasidiabatic and
adiabatic Hamiltonians are exact unitary transformations of each
other.\cite{Choi_Vanicek:2020}

The exact quasidiabatic Hamiltonian from Eq.~(\ref{eq:mol_H_accurate}) cannot
be expressed as a sum of terms depending purely on either the position or
momentum operator. Due to this nonseparable nature of the Hamiltonian, we
require an integrator that is applicable to any form of the Hamiltonian. For
example, the popular split-operator algorithm\cite{Feit_Steiger:1982,
book_Lubich:2008, book_Tannor:2007, Roulet_Vanicek:2019} cannot be employed.
The wavepackets are, therefore, propagated with the
composition\cite{Suzuki:1990, Yoshida:1990, book_Hairer_Wanner:2006,
book_Lubich:2008} of the implicit midpoint
method,\cite{book_Hairer_Wanner:2006, book_Leimkuhler_Reich:2004,
McCullough_Wyatt:1971} which, like the closely related trapezoidal rule (or
Crank--Nicolson method)%
,\cite{Crank_Nicolson:1947,McCullough_Wyatt:1971} works for both separable and
nonseparable Hamiltonians, as long as the action of the Hamiltonian on the
wavepacket can be evaluated. Moreover, in contrast to some other
methods applicable to nonseparable Hamiltonians, the chosen methods preserve
exactly most geometric properties of the exact solution: conservation of the
norm, energy, and inner-product, linearity, symplecticity, stability,
symmetry, and time reversibility.\cite{Choi_Vanicek:2019}

For a valid comparison of the two wavepackets propagated with either the exact
or approximate quasidiabatic Hamiltonian, the numerical errors must be much
smaller than the errors due to omitting the residual couplings. Owing to its
exact symmetry, the implicit midpoint method can be composed using various
schemes\cite{Suzuki:1990, Yoshida:1990, Kahan_Li:1997,
book_Hairer_Wanner:2006} to obtain integrators of arbitrary even orders of
accuracy in the time step;\cite{Choi_Vanicek:2019, Roulet_Vanicek:2019} we
compose the implicit midpoint method according to the optimal
scheme\cite{Kahan_Li:1997} to obtain an eighth-order integrator. By using this
high-order integrator with a small time step, the time discretization errors
are kept negligible (see Sec.~S2 of the supplementary material).

\section{\label{sec:numerical_examples}Numerical examples}

We now apply the method proposed in Sec.~\ref{sec:theory} to
nonadiabatic quantum simulations in the cubic $E\otimes e$ Jahn--Teller model
of NO$_{3}$\cite{book_Bersuker_Polinger:2012, Viel_Eisfeld:2004,
Bersuker:2001, Hauser_Ernst:2010} and in the induced Renner--Teller model of
HCN.\cite{Pacher_Koppel:1988, Koppel_Von_Niessen:1979, Cederbaum_Domcke:1981,
Koppel_Cederbaum:1981} Despite their reduced dimensionality, these two-dimensional ($D=2$), two-state ($S=2$) models exhibit interesting dynamics\cite{book_Domcke_Koppel:2004, Viel_Eisfeld:2004, Koppel_Cederbaum:1981} due to the presence of strong nonadiabatic couplings; in particular, $\mathbf{F}_{\mathrm{ad}}(Q)$ diverges at $Q=0$, the point of intersection between the two adiabatic potential energy surfaces.  

In both models, doubly degenerate electronic states labeled by $n=1$ and $n=2$
are coupled by doubly degenerate normal modes $Q_{1}$ and $Q_{2}$. We use
\textquotedblleft natural\textquotedblright\ units (n.u.) throughout by
setting $k=M=\hbar=1$ n.u., where $M$ is the mass associated with the
degenerate normal modes (which differs from the electron mass used
in atomic units), and $\hbar\omega=\hbar\sqrt{k/M}$ is the quantum of the
vibrational energy of these modes. Whenever convenient, we express
the potential energy surface in polar coordinates: the radius $\rho
(Q):=\sqrt{Q_{1}^{2}+Q_{2}^{2}}$ and polar angle $\phi(Q):=\arctan
{(Q_{2}/Q_{1})}$.

All numerical wavepacket propagations were performed with a small time step of
$\Delta t=1/(40\omega)=0.025$ n.u. on a uniform grid of $N\times N$ points
defined between $Q_{l}=-Q_{\mathrm{lim}}$ and $Q_{l}=Q_{\mathrm{lim}}$ in both
nuclear dimensions: $N$ = 64 and $Q_{\mathrm{lim}}=10$ n.u. in
the NO$_{3}$ model, while $N$ = 32 and $Q_{\mathrm{lim}}=7$ n.u. in the HCN
model.

\subsection{\label{sec:no3_vert}Jahn--Teller effect in nitrogen trioxide}

Although the strictly diabatic Hamiltonian
\begin{equation}
\hat{\mathbf{H}}_{\mathrm{diab}}=\frac{\hat{P}^{2}}{2M}\mathbf{1}%
+\mathbf{V}_{\mathrm{diab}}(\hat{Q})\label{eq:jt_H_ref}%
\end{equation}
does not exist in general, it may exist exceptionally and, in fact, is used to
define the Jahn--Teller model.\cite{book_Bersuker_Polinger:2012,
Viel_Eisfeld:2004, Bersuker:2001, Hauser_Ernst:2010, Thiel_Koppel:1999} In
Eq.~(\ref{eq:jt_H_ref}), the diabatic potential energy matrix 

\begin{equation}
\mathbf{V}_{\mathrm{diab}}(Q)=%
\begin{pmatrix}
E_{0}(Q) & E_{\mathrm{cpl}}(Q)\\
E_{\mathrm{cpl}}(Q)^{\ast} & E_{0}(Q)
\end{pmatrix}
\end{equation}
depends on the cubic potential energy $E_{0}(\rho,\phi):=k\rho
^{2}/2+2\alpha\rho^{3}\cos{3\phi}$ and Jahn--Teller
coupling\cite{Viel_Eisfeld:2004}
\begin{equation}
E_{\mathrm{cpl}}(\rho,\phi):=f(\rho)e^{-i\phi}+c_{2}\rho^{2}e^{2i\phi
},\label{eq:JT_E_cpl}%
\end{equation}
where $f(\rho):=c_{1}\rho+c_{3}\rho^{3}$. In our nonadiabatic simulations
of nitrogen trioxide, we used the Jahn--Teller model of NO$_{3}$ from
Ref.~\onlinecite{Viel_Eisfeld:2004} with parameters $\alpha=-0.0125$ n.u.,
$c_{1}=0.375$ n.u., $c_{2}=-0.0668$ n.u., and $c_{3}=-0.0119$ n.u. To simplify
the following presentation, we rewrite $E_{\mathrm{cpl}}(Q)$ as
$E_{\mathrm{cpl}}(Q)=|E_{\mathrm{cpl}}(Q)|e^{-2i\theta(Q)}$ using the mixing
angle
\begin{equation}
\theta(\rho,\phi):=\frac{1}{2}\arctan{\frac{f(\rho)\sin{\phi}-c_{2}\rho
^{2}\sin{2\phi}}{f(\rho)\cos{\phi}+c_{2}\rho^{2}\cos{2\phi}}}%
.\label{eq:JT_mixing_angle}%
\end{equation}

Our previous study\cite{Choi_Vanicek:2020} on a similar system showed that the
exact quasidiabatic and strictly diabatic Hamiltonians yield nearly identical
results. Here, however, we intentionally avoid using the strictly
diabatic Hamiltonian as a benchmark and use it only to define the model, in
order that the approach and conclusions of this study are applicable also to
systems where the strictly diabatic Hamiltonian does not
exist\cite{Mead_Truhlar:1982, Pacher_Koppel:1989} (see Sec.~\ref{sec:hcn}
below for an explicit example of such a system).

\begin{figure}
[tbp]\includegraphics[]{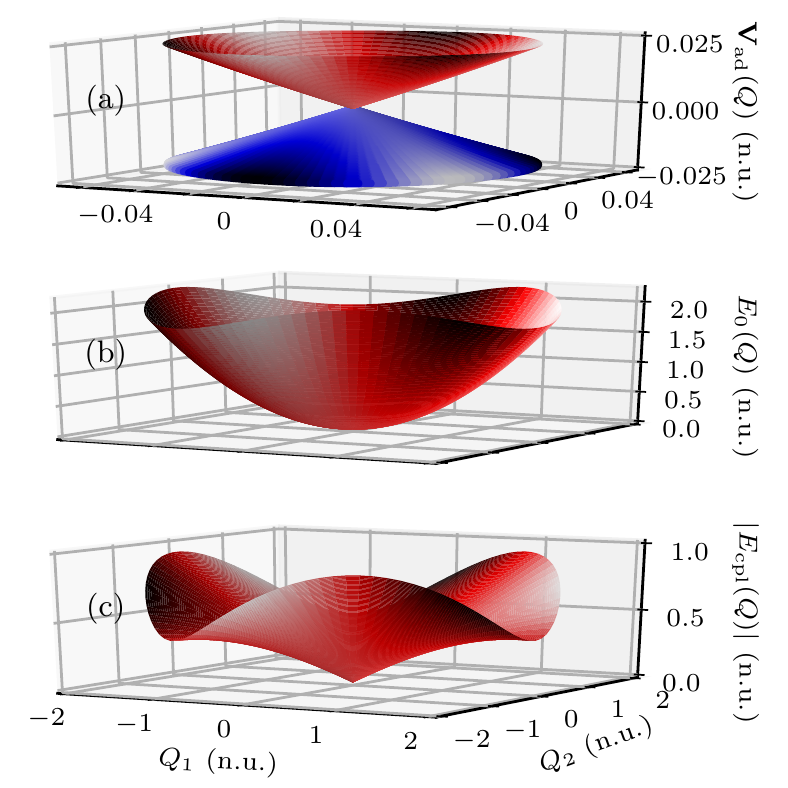}
\caption{Potential energy surfaces in the cubic $E \otimes e$ model of the Jahn--Teller effect in NO$_{3}$
in the vicinity of the conical intersection at $Q = 0$. (a) Elements $V_{1}(Q) = V_{+}(Q)$ (red) and
$V_{2}(Q) = V_{-}(Q)$ (blue) of the diagonal adiabatic potential energy matrix;
the two surfaces intersect (touch) at the point $Q=0$. The diabatic potential energy matrix consists of (b) the cubic potential energy surfaces $E_{0}(Q)$ on the diagonal and (c) the off-diagonal complex couplings of magnitude $|E_{\mathrm{cpl}}(Q)|$.}\label{fig:no3_pes}%

\end{figure}

The adiabatic states in the Jahn--Teller model are obtained by a process
inverse to diabatization, i.e., by a unitary transformation of the strictly
diabatic states using any matrix that diagonalizes $\mathbf{V}_{\mathrm{diab}%
}(Q)$. Following
Refs.~\onlinecite{book_Bersuker_Polinger:2012, Thiel_Koppel:1999}, we employed
the transformation matrix
\begin{equation}
\mathbf{T}(Q)=\frac{1}{\sqrt{2}}%
\begin{pmatrix}
e^{-i\theta(Q)} & e^{-i\theta(Q)}\\
e^{i\theta(Q)} & -e^{i\theta(Q)}%
\end{pmatrix}
.\label{eq:JT_transf_matrix}%
\end{equation}
In the resulting adiabatic representation, the diagonal potential energy
matrix has elements $V_{1}(Q)=V_{+}(Q)$ and $V_{2}(Q)=V_{-}(Q)$, where
$V_{\pm}(Q):=E_{0}(Q)\pm|E_{\mathrm{cpl}}(Q)|$. [Matrix elements of
$\mathbf{V}_{\mathrm{ad}}(Q)$ and $\mathbf{V}_{\mathrm{diab}}(Q)$ are plotted
in Fig.~\ref{fig:no3_pes}.] Transformation~(\ref{eq:JT_transf_matrix}) also
yields analytical expressions for the nonadiabatic vector
couplings\cite{Viel_Eisfeld:2004, book_Bersuker_Polinger:2012}
\begin{equation}
\mathbf{F}_{\mathrm{ad}}(Q)=-i\nabla\theta(Q)%
\begin{pmatrix}
0 & 1\\
1 & 0
\end{pmatrix}
\end{equation}
and for the nonadiabatic scalar couplings
\begin{equation}
\mathbf{G}_{\mathrm{ad}}(Q)=-%
\begin{pmatrix}
\lbrack\nabla\theta(Q)]^{2} & i\nabla^{2}\theta(Q)\\
i\nabla^{2}\theta(Q) & [\nabla\theta(Q)]^{2}%
\end{pmatrix}
.
\end{equation}
As expected, the nonadiabatic couplings diverge at the conical intersection at
$\rho=0$ since the azimuthal component of $\mathbf{F}_{\mathrm{ad}}(Q)$ is
proportional to 
\begin{equation}
\rho^{-1}\frac{\partial\theta(\rho,\phi)}{\partial\phi}=\frac{f(\rho)^{2}%
-2c_{2}^{2}\rho^{4}-c_{2} \rho^{2} f(\rho) \cos{3\phi}}{2\rho|E_{\mathrm{cpl}%
}(\rho, \phi)|^{2}}\xrightarrow{\rho \to 0}\infty.\label{eq:res_coupling_phi}%
\end{equation}

In the cubic Jahn--Teller model, the regularized diabatization
scheme\cite{Thiel_Koppel:1999, Koppel_Mahapatra:2001, Koppel_Schubert:2006}
can be implemented analytically. The $j$th-order adiabatic to quasidiabatic
transformation matrix
\begin{equation}
\mathbf{S}(Q)=\frac{1}{\sqrt{2}}%
\begin{pmatrix}
e^{-i\theta^{(j)}(Q)} & e^{-i\theta^{(j)}(Q)}\\
e^{i\theta^{(j)}(Q)} & -e^{i\theta^{(j)}(Q)}%
\end{pmatrix}
\label{eq:JT_S_matrix}%
\end{equation}
is obtained simply by replacing $\theta(Q)$ with $\theta^{(j)}(Q)$ in
Eq.~(\ref{eq:JT_transf_matrix}) for $\mathbf{T}(Q)$: $\theta^{(1)}%
(\rho,\phi):=\phi/2$ in the first-order scheme,
\begin{equation}
\theta^{(2)}(\rho,\phi):=\frac{1}{2}\arctan{\frac{c_{1}\rho\sin{\phi}%
-c_{2}\rho^{2}\sin{2\phi}}{c_{1}\rho\cos{\phi}+c_{2}\rho^{2}\cos{2\phi}}%
}\label{eq:theta_2}%
\end{equation}
in the second-order scheme, while---in the cubic Jahn--Teller model---the
third-order quasidiabatization is already identical to the strict
diabatization, i.e., $\theta^{(3)}(\rho,\phi)=\theta(\rho,\phi)$. The
quasidiabatization yields the potential energy matrix
\begin{align}
\mathbf{V}_{\mathrm{qd}}(Q) &  :=\mathbf{S}(Q)\mathbf{V}_{\mathrm{ad}%
}(Q)\mathbf{S}(Q)^{\dagger}\nonumber\\
&  =%
\begin{pmatrix}
E_{0}(Q) & |E_{\mathrm{cpl}}(Q)|e^{-2i\theta^{(j)}(Q)}\\
|E_{\mathrm{cpl}}(Q)|e^{2i\theta^{(j)}(Q)} & E_{0}(Q)
\end{pmatrix}
,\label{eq:V_qd}%
\end{align}
residual vector couplings
\begin{align}
\mathbf{F}_{\mathrm{qd}}(Q) &  :=\mathbf{S}(Q)\mathbf{F}_{\mathrm{ad}%
}(Q)\mathbf{S}(Q)^{\dagger}+\mathbf{S}(Q)\nabla\mathbf{S}(Q)^{\dagger
}\nonumber\\
&  =-i\nabla\theta_{-}^{(j)}(Q)%
\begin{pmatrix}
1 & 0\\
0 & -1
\end{pmatrix}
,
\end{align}
and residual scalar couplings
\begin{align}
\mathbf{G}_{\mathrm{qd}}(Q) &  :=\mathbf{S}(Q)\mathbf{G}_{\mathrm{ad}%
}(Q)\mathbf{S}(Q)^{\dagger}\nonumber\\
+2 &  \mathbf{S}(Q)\mathbf{F}_{\mathrm{ad}}(Q)\nabla\mathbf{S}(Q)^{\dagger
}+\mathbf{S}(Q)\nabla^{2}\mathbf{S}(Q)^{\dagger}\nonumber\\
&  =-i\nabla^{2}\theta_{-}^{(j)}(Q)%
\begin{pmatrix}
1 & 0\\
0 & -1
\end{pmatrix}
-[\nabla\theta_{-}^{(j)}(Q)]^{2}%
\begin{pmatrix}
1 & 0\\
0 & 1
\end{pmatrix}
,\label{eq:JT_residual_scalar_couplings}%
\end{align}
where $\theta_{-}^{(j)}(Q):=\theta(Q)-\theta^{(j)}(Q)$. The resulting
magnitude of the residual couplings in the first-order ($j=1$) scheme is
$\mathcal{R}[\mathbf{F}_{\mathrm{qd}}(Q)]=3.8$ n.u.

The hermiticity of Hamiltonian~(\ref{eq:mol_H_accurate}) is broken on a finite
grid because the commutator relation $[\hat{P},\mathbf{F}_{\mathrm{qd}}%
(\hat{Q})]=-i\hbar\nabla\cdot\mathbf{F}_{\mathrm{qd}}(\hat{Q})$ holds only
approximately unless the grid is infinitely dense. Yet, the hermiticity of the
Hamiltonian is essential for the norm conservation (see Fig.~S5 in
Sec.~S3 of the supplementary material), which, in turn, is required for
quantum fidelity $\mathcal{F}(t)$ and distance $\mathcal{D}(t)$ to be valid
measures of the importance of the residual nonadiabatic couplings. To make the
exact quasidiabatic Hamiltonian exactly Hermitian, we re-express it as
\begin{equation}
\hat{\mathbf{H}}_{\text{qd-exact}}=\frac{1}{2M}[\hat{P}\mathbf{1}%
-i\hbar\mathbf{F}_{\mathrm{qd}}(\hat{Q})]^{2}+\mathbf{V}_{\mathrm{qd}}(\hat
{Q}),\label{eq:mol_H_qdiab_simple}%
\end{equation}
using the relationship
\begin{equation}
\mathbf{G}_{\mathrm{qd}}(Q)=\nabla\cdot\mathbf{F}_{\mathrm{qd}}(Q)+\mathbf{F}%
_{\mathrm{qd}}(Q)^{2},\label{eq:nonad_scalar_coupling}%
\end{equation}
which holds---exceptionally---for systems, such as the Jahn--Teller model,
that can be represented exactly by a finite number of states; in general,
Eq.~(\ref{eq:nonad_scalar_coupling}) only holds when $S\rightarrow\infty$.

Another benefit of Hamiltonian~(\ref{eq:mol_H_qdiab_simple}) is the absence of $\mathbf{G}_{\mathrm{qd}}(Q)$, the evaluation of which represents the computational bottleneck in realistic systems. Likewise, the equations of motion in the widely-employed Meyer--Miller approach\cite{Miller_McCurdy:1978, Meyer_Miller:1979, Meyer_Miller:1980, Stock_Thoss:1997} can be simplified greatly\cite{Cotton_Miller:2017} by starting from Hamiltonian~(\ref{eq:mol_H_qdiab_simple}) instead of Hamiltonian~(\ref{eq:mol_H_accurate}). These two forms of the molecular Hamiltonian, however, are strictly equivalent only if the electronic basis is complete.\cite{Yarkony:1996, Pacher_Koppel:1989} In generic systems, in which the relation~(\ref{eq:nonad_scalar_coupling}) does not hold and one is obliged to use the original Hamiltonian~(\ref{eq:mol_H_accurate}) and evaluate $\mathbf{G}_{\mathrm{qd}}(Q)$, the computationally expensive evaluation of the second derivatives of electronic wavefunctions with respect to nuclear coordinates can still be avoided by using the relation 
\begin{equation}
\mathbf{G}_{\mathrm{qd}}(Q) = \nabla \cdot \mathbf{F}_{\mathrm{qd}}(Q) - \mathbf{K}_{\mathrm{qd}}(Q),\label{eq:nonad_scalar_coupling_gen}
\end{equation}
where $[\mathbf{K}_{\mathrm{qd}}(Q)]_{mn} := \langle \nabla m^{\prime}(Q) | \nabla n^{\prime}(Q) \rangle$ requires only the first derivatives of the quasidiabatic electronic states $|n^{\prime}(Q)\rangle$, introduced in Eq.~(\ref{eq:qd_el_states}). In contrast to Eq.~(\ref{eq:nonad_scalar_coupling}), relation~(\ref{eq:nonad_scalar_coupling_gen}) holds in arbitrary systems and for finite $S$.

To analyze the importance of residual couplings in NO$_{3}$, we simulated,
with either the exact or approximate quasidiabatic Hamiltonian, the quantum
dynamics following an electronic transition from the ground vibrational
eigenstate of the ground electronic state $V_{\mathrm{g}}(Q)=-E_{\mathrm{gap}%
}+k\rho(Q)^{2}/2$ with $E_{\mathrm{gap}}=11$ n.u. ($1$ n.u. of
energy here corresponds to $0.2$ eV~$\approx0.007$ a.u.). Invoking the
time-dependent perturbation theory and Condon approximation, we considered the
initial state in the quasidiabatic representation to be 
\begin{equation}
\bm{\psi}(Q,t=0):=\frac{e^{-\rho(Q)^{2}/2\hbar}}{\sqrt{2\pi\hbar}}%
\begin{pmatrix}
1\\
1
\end{pmatrix}
,\label{eq:init_state}%
\end{equation}
where we omitted (and will omit) the superscript \textquotedblleft
qd\textquotedblright\ on the wavepacket for brevity.

\begin{figure}
[tbp]\includegraphics[]{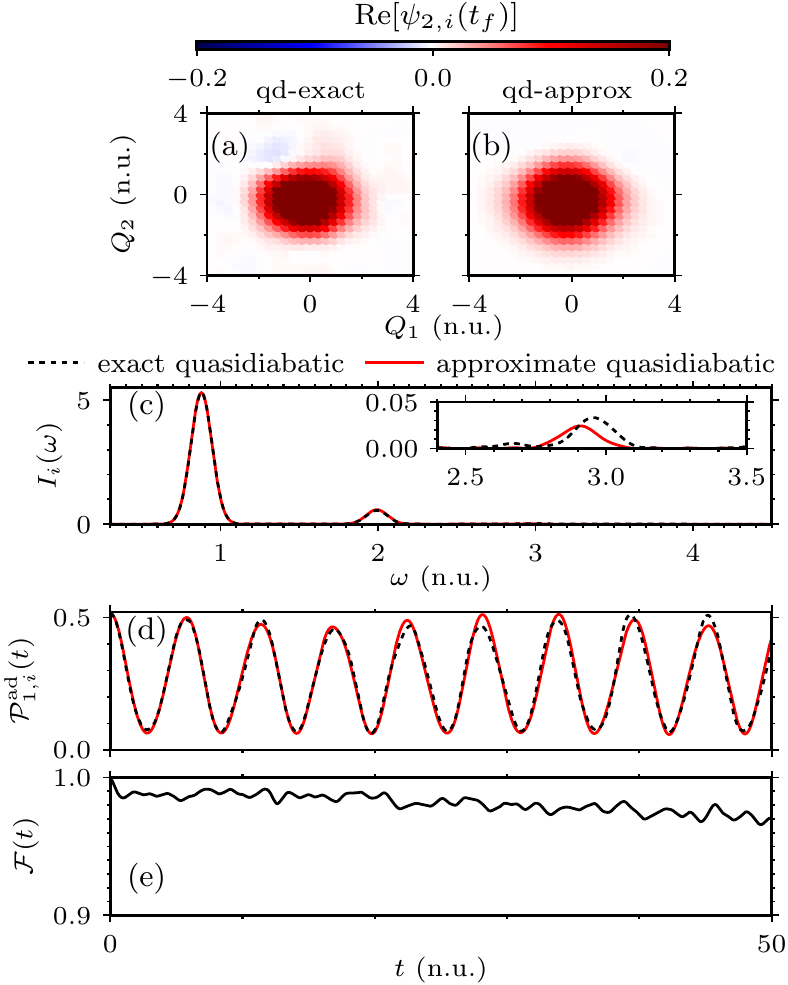}
\caption{Importance of the residual nonadiabatic couplings in the NO$_{3}$ model from
Sec.~\ref{sec:no3_vert}.
The figure compares the wavepackets and observables obtained with either the
exact ($i=\textrm{qd-exact}$) or approximate ($i=\textrm{qd-approx}$) quasidiabatic Hamiltonian.
(a) and (b): Wavepackets propagated with (a)  $\hat{\mathbf{H}}_{\text{qd-exact}}$ from
Eq.~(\ref{eq:mol_H_accurate}) and (b)  $\hat{\mathbf{H}}_{\text{qd-approx}}$ from Eq.~(\ref{eq:mol_H_approx}).
[Only the real part of the nuclear
wavepacket in the second ($n=2$) electronic state is shown.]
(c) Power spectrum $I_{i}(\omega)$ obtained by Fourier transforming the damped autocorrelation function.
[To emulate the broadening of the spectral peaks, the autocorrelation function
$C_{i}(t) = \langle \bm{\psi}(0) | \bm{\psi}_{i}(t) \rangle$ was multiplied by the damping function
$d(t) = \exp[(-t/t_{\mathrm{damp}})^{2}]$ with $t_{\mathrm{damp}} = 17.5$ n.u.]
(d) Population $\mathcal{P}_{1,i}^{\mathrm{ad}}(t) := \langle \bm{\psi}_{i}^{\mathrm{ad}}(t) | \mathbf{P}_{1} |
\bm{\psi}_{i}^{\mathrm{ad}}(t) \rangle$ of the first ($n=1$) adiabatic electronic state;
$\mathbf{P}_{n} := | n \rangle \langle n |$ is the population operator
of the $n$th adiabatic state.
(e) Errors due to ignoring the residual couplings are measured by quantum fidelity $\mathcal{F}(t)$
[Eq.~(\ref{eq:fidelity})].}\label{fig:no3_vert_main}
\end{figure}

Figure \ref{fig:no3_vert_main} shows that, in the nonadiabatic
dynamics following the vertical excitation of NO$_{3}$, neglecting the
residual couplings does not significantly affect the wavepacket [compare
panels (a) and (b)], power spectrum $I(\omega)$ [panel~(c)], or population
$\mathcal{P}_{1}^{\mathrm{ad}}(t)$ [panel~(d)]. Even the fidelity
$\mathcal{F}(t)$ [panel~(e)] between the wavepackets propagated either with or
without the residual couplings remains close to the maximal value of $1$ until
the final time $t_{f}$. Section~S4 of the supplementary material
further supports this conclusion by displaying the time dependence of position
$\langle\rho\rangle(t)$, potential energy $\langle\mathbf{V}_{\mathrm{qd}%
}\rangle(t)$, and distance $\mathcal{D}(t)$. In contrast, as we will see in
Secs.~\ref{sec:hcn} and \ref{sec:no3_disp}, the residual couplings are much
more significant in the HCN model and in the displaced excitation of NO$_{3}%
$.

In Sec.~S5 of the supplementary material, we also analyze the
importance of the residual couplings for different Jahn--Teller coupling
coefficients and different initial populations.

\subsection{\label{sec:hcn}Induced Renner--Teller effect in
hydrogen cyanide}

The model of the induced Renner--Teller effect\cite{Pacher_Koppel:1988,
Koppel_Von_Niessen:1979, Cederbaum_Domcke:1981, Koppel_Cederbaum:1981} is more
realistic than the Jahn--Teller model from Sec.~\ref{sec:no3_vert}: In
particular, the strictly diabatic Hamiltonian~(\ref{eq:jt_H_ref}) cannot be
defined and relationship~(\ref{eq:nonad_scalar_coupling}) does not hold.
Nevertheless, similarly to the Jahn--Teller model, the nonadiabatic couplings
between the adiabatic states are singular at $Q=0$.\cite{Pacher_Koppel:1988}
Since Eq.~(\ref{eq:nonad_scalar_coupling}) does not hold in the induced
Renner--Teller model, the exactly Hermitian
Hamiltonian~(\ref{eq:mol_H_qdiab_simple}) cannot be used instead of
Hamiltonian~(\ref{eq:mol_H_accurate}). Yet, even with
Hamiltonian~(\ref{eq:mol_H_accurate}), the norm is sufficiently converged in
grid density for the quantum fidelity $\mathcal{F}(t)$ and distance
$\mathcal{D}(t)$ to be valid (see Fig.~S6 of the supplementary material).

\begin{figure}
[tbp]\includegraphics[]{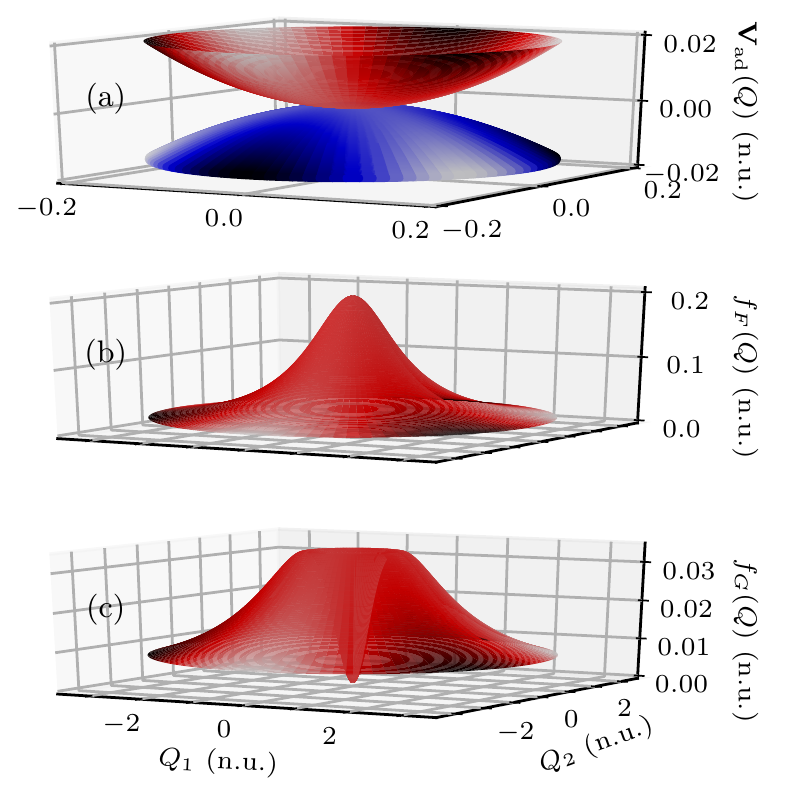}
\caption{Potential energy surfaces in the model of the induced Renner--Teller effect in HCN in the vicinity
of the Renner--Teller
intersection at $Q = 0$. (a) The two adiabatic potential energy surfaces $V_{1}(Q)$ (blue) and $V_{2}(Q)$ (red)
intersect (touch) at the point $Q=0$. The residual couplings~(\ref{eq:rt_F}) and (\ref{eq:rt_G}) depend on
plotted functions $f_{F}(Q)$ [panel (b)] and $f_{G}(Q)$ [panel (c)].}\label{fig:hcn_pes}%

\end{figure}

We follow Ref.~\onlinecite{Pacher_Koppel:1988}, where the induced
Renner--Teller model is quasidiabatized with the block-diagonalization scheme,
which minimizes the residual couplings locally (around $Q=0$ in this
model).\cite{Pacher_Koppel:1989} The resulting quasidiabatic potential energy
matrix
\begin{equation}
\mathbf{V}_{\mathrm{qd}}(\rho,\phi)=%
\begin{pmatrix}
V_{+}(\rho) & V_{-}(\rho)e^{-2i\phi}\\
V_{-}(\rho)e^{2i\phi} & V_{+}(\rho)
\end{pmatrix}
\label{eq:rt_qd_pot}%
\end{equation}
with $V_{\pm}(\rho):=[V_{1}(\rho)\pm V_{2}(\rho)]/2$ depends on the
adiabatic potential energy surfaces $V_{1}(\rho)=\Delta+E_{\mathrm{h}%
}(\rho)-w(\rho)$ and $V_{2}(\rho)=E_{\mathrm{h}}(\rho)$, where
$E_{\mathrm{h}}(\rho):=k\rho^{2}/2$ and $w(\rho):=(\Delta^{2}+2\lambda
^{2}\rho^{2})^{1/2}$. Analytical expressions for the nonadiabatic couplings,
$\mathbf{F}_{\mathrm{ad}}(Q)$ and $\mathbf{G}_{\mathrm{ad}}(Q)$, and adiabatic
to quasidiabatic transformation matrix $\mathbf{S}(Q)$ can be found in
Ref.~\onlinecite{Pacher_Koppel:1988}. In our nonadiabatic simulations of
hydrogen cyanide, we used the induced Renner--Teller model of HCN from
Refs.~\onlinecite{Pacher_Koppel:1988, Koppel_Von_Niessen:1979} with parameters
$\Delta=1.11$ n.u. and $\lambda=1$ n.u. The residual vector and scalar
couplings\cite{Pacher_Koppel:1988} are, respectively, \
\begin{equation}
\mathbf{F}_{\mathrm{qd}}(Q)=iv_{F}(Q)%
\begin{pmatrix}
1 & 0\\
0 & -1
\end{pmatrix}
\label{eq:rt_F}%
\end{equation}
and
\begin{equation}
\mathbf{G}_{\mathrm{qd}}(\rho,\phi)=2%
\begin{pmatrix}
f_{F}(\rho)-f_{G}(\rho) & -f_{G}(\rho)e^{-2i\phi}\\
-f_{G}(\rho)e^{2i\phi} & f_{F}(\rho)-f_{G}(\rho)
\end{pmatrix}
,\label{eq:rt_G}%
\end{equation}
where we have defined the $D$-dimensional (here $D=2$) vector $v_{F}(Q)$ with
components $v_{F}(Q)_{1}=-f_{F}(\rho(Q))Q_{2}$ and $v_{F}(Q)_{2}%
=f_{F}(\rho(Q))Q_{1},$ and functions $f_{F}(\rho):=[1-w_{+}(\rho
)]/\rho^{2}$ and
\begin{equation}
f_{G}(\rho):=\left[  \frac{w_{-}(\rho)}{2\rho}\right]  ^{2}-\left[
\frac{\lambda}{2\sqrt{2}w(\rho)}\right]  ^{2}+\left[  \frac{\lambda^{2}\rho
}{2w(\rho)^{2}}\right]  ^{2}%
\end{equation}
with $w_{\pm}(\rho):=\sqrt{[1\pm\Delta/w(\rho)]/2}$. The magnitude of the
residual couplings~(\ref{eq:rt_F}) is $\mathcal{R}[\mathbf{F}_{\mathrm{qd}%
}(Q)]=0.37$ n.u.; the adiabatic potential energy surfaces and functions
$f_{F}$ and $f_{G}$ are plotted in Fig.~\ref{fig:hcn_pes}. 

\begin{figure}
[tbp]\includegraphics[]{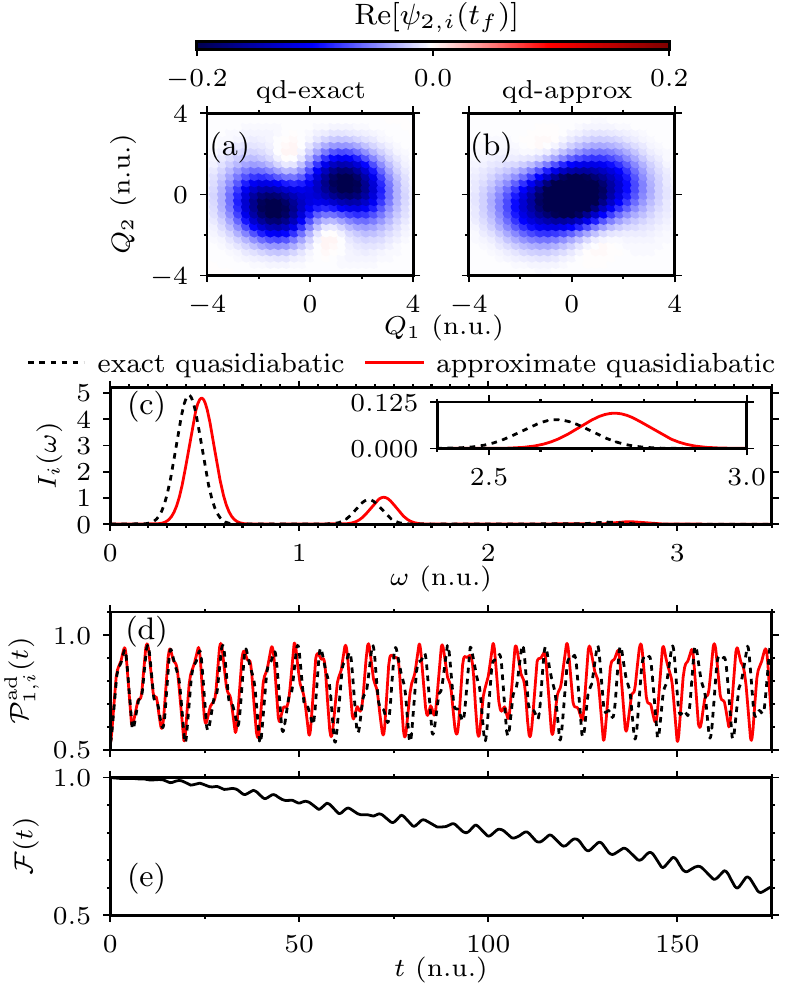}
\caption{Importance of the residual nonadiabatic couplings in the HCN model from Sec.~\ref{sec:hcn}.
(a) and (b): wavepackets, (c) power spectrum, (d) population, (e) fidelity. See the caption of Fig.~\ref{fig:no3_vert_main} for a detailed description of the content of the five panels.}\label{fig:hcn_main}%

\end{figure}

Similarly to Sec.~\ref{sec:no3_vert}, we simulate the dynamics following an
electronic transition from the ground vibrational eigenstate of
$V_{\mathrm{g}}(\rho,\phi)=-E_{\mathrm{gap}}+E_{\mathrm{h}}(\rho)$ with
$E_{\mathrm{gap}}=153$ n.u. ($1$ n.u. of energy here corresponds
to $0.09$ eV~$\approx0.003$ a.u.). Unlike their analogues in
Sec.~\ref{sec:no3_vert}, however, the two wavepackets, propagated with either
the exact or approximate quasidiabatic Hamiltonian, differ significantly
[compare panels (a) and (b) of Fig.~\ref{fig:hcn_main}]. Ignoring the residual
couplings also leads to large errors in the power spectrum $I(\omega)$
[panel~(c)], population $\mathcal{P}_{1}^{\mathrm{ad}}(t)$ [panel~(d)], and
fast decay of quantum fidelity $\mathcal{F}(t)$ [panel~(e)]. In particular,
the population obtained with the approximate quasidiabatic Hamiltonian cannot
be trusted because, e.g., at $t=169$ n.u., the error $\epsilon_{\text{res-cpl}%
}[\mathcal{P}_{1}^{\mathrm{ad}}(t)]:=|\mathcal{P}_{1,\text{qd-approx}%
}^{\mathrm{ad}}(t)-\mathcal{P}_{1,\text{qd-exact}}^{\mathrm{ad}}(t)|$ due to
the neglect of the residual couplings is of the same order as the range
$R_{\mathcal{P}_{1}^{\mathrm{ad}}}:=\mathcal{P}_{1,\mathrm{max}}^{\mathrm{ad}%
}-\mathcal{P}_{1,\mathrm{min}}^{\mathrm{ad}}$ of the population in the whole
simulation interval: $\epsilon_{\text{res-cpl}}[\mathcal{P}_{1}^{\mathrm{ad}%
}(t)]/R_{\mathcal{P}_{1}^{\mathrm{ad}}}=0.7$. Note also that neither
$\mathcal{P}_{1}^{\mathrm{ad}}(t)$ nor $\mathcal{F}(t)$ is affected by the
overall phases of the two wavepackets, although a linearly growing overall
phase difference appears to be the main contribution to the change of the
spectrum, which is mostly shifted [see panel (c) of Fig.~\ref{fig:hcn_main}].

\subsection{\label{sec:no3_disp}Displaced excitation of nitrogen trioxide}

Although the excited states of NO$_{3}$ from
Sec.~\ref{sec:no3_vert} are bright states, the coupled states modeled by the
cubic $E\otimes e$ Jahn--Teller Hamiltonian can sometimes be dark. A
wavepacket might reach such dark states at a nuclear geometry that is not the
ground state equilibrium (e.g., via an intersection with a bright state).
Motivated by this observation from Ref.~\onlinecite{Viel_Eisfeld:2004}, we
consider the initial state~(\ref{eq:init_state}) displaced in both $Q_{1}$ and
$Q_{2}$ by $-0.8$ n.u. Keeping all other parameters fixed as in
Sec.~\ref{sec:no3_vert} will allow us to analyze how the
importance of the residual couplings depends on the initial state and on the
quasidiabatization method used.

\begin{figure}
[tbp]\includegraphics[]{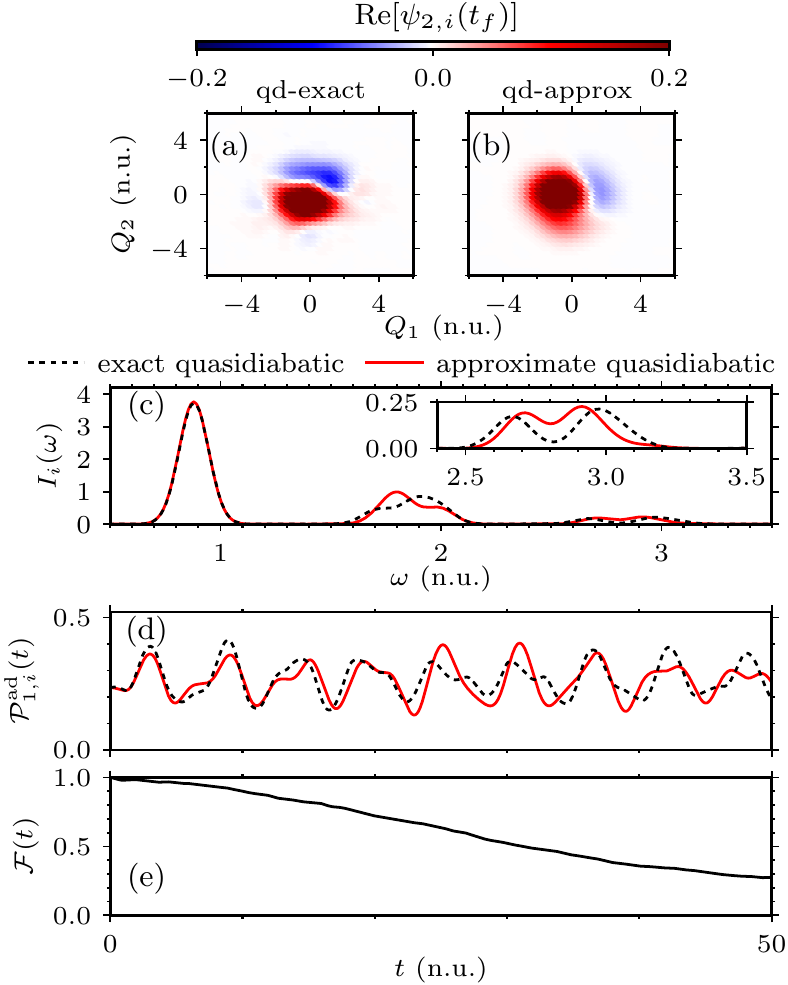}
\caption{Importance of the residual nonadiabatic couplings in the model of a displaced excitation of
NO$_{3}$ from Sec~\ref{sec:no3_disp}. As in Fig.~\ref{fig:no3_vert_main}, the molecular Hamiltonian was
quasidiabatized with the first-order ($j=1$) scheme. (a) and (b): wavepackets, (c) power spectrum, (d) population, (e) fidelity. See the caption of Fig.~\ref{fig:no3_vert_main} for a detailed description of the content of the five panels.
} \label{fig:no3_disp_main_1}
\end{figure}

Figure~\ref{fig:no3_disp_main_1} shows that, in contrast to the vertical
excitation from Sec.~\ref{sec:no3_vert} (analyzed in
Fig.~\ref{fig:no3_vert_main}), ignoring the residual couplings obtained by
applying the first-order regularized scheme to the displaced
excitation of NO$_{3}$ affects both the wavepacket [compare panels (a) and
(b)] and observables significantly. Neither the spectrum $I(\omega)$ [panel
(c)] nor population $\mathcal{P}_{1}^{\mathrm{ad}}(t)$ [panel (d)] is obtained
accurately with the approximate quasidiabatic Hamiltonian. For example, at
$t=25$ n.u., the error of the population is almost half of the range of the
population in the whole simulation interval: $\epsilon_{\text{res-cpl}%
}[\mathcal{P}_{1}^{\mathrm{ad}}(t)]/R_{\mathcal{P}_{1}^{\mathrm{ad}}}=0.4$.
The quantum fidelity [panel (e)] decreases rapidly to $\mathcal{F}%
(t_{f})\approx0.3$ at $t_{f}=50$ n.u.

\begin{figure}
[tbp]\includegraphics[]{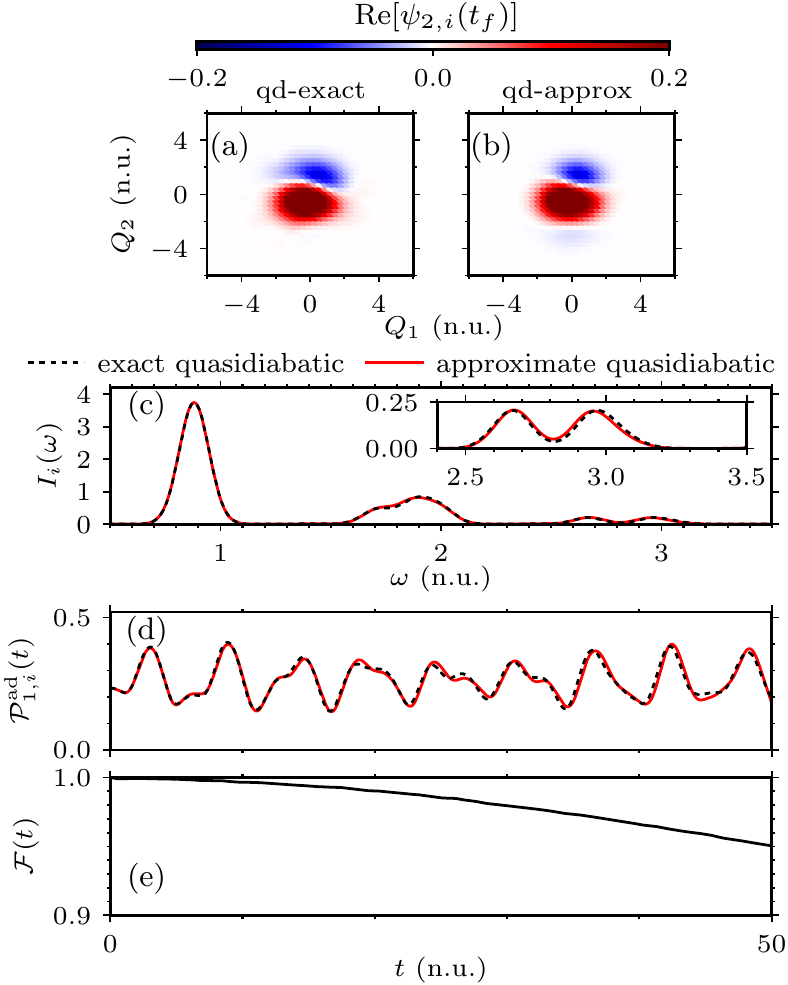}
\caption{Importance of the residual nonadiabatic couplings in the model of a displaced excitation of
NO$_{3}$ from Sec~\ref{sec:no3_disp}.  The only difference from
Fig.~\ref{fig:no3_disp_main_1} is that the molecular Hamiltonian was
quasidiabatized with the second-order ($j=2$) scheme. (a) and (b): wavepackets, (c) power spectrum, (d) population, (e) fidelity. See the caption of Fig.~\ref{fig:no3_vert_main} for a detailed description of the content of the five panels.
} \label{fig:no3_disp_main_2}
\end{figure}

The residual couplings, however, can be made less important by an improved
quasidiabatization. One can reduce the magnitude of the residual couplings
from $\mathcal{R}[\mathbf{F}_{\mathrm{qd}}^{(1)}(Q)]=3.8$ n.u. to
$\mathcal{R}[\mathbf{F}_{\mathrm{qd}}^{(2)}(Q)]=0.5$ n.u. by employing the
more sophisticated, second-order regularized diabatization
scheme\cite{Thiel_Koppel:1999, Koppel_Mahapatra:2001, Koppel_Schubert:2006}
obtained by inserting $\theta^{(2)}(Q)$ from Eq.~(\ref{eq:theta_2}) in
Eqs.~(\ref{eq:JT_S_matrix}) and (\ref{eq:V_qd}%
)--(\ref{eq:JT_residual_scalar_couplings}). When this second-order scheme is
used, the errors of the wavepacket $\bm{\psi}(t)$, spectrum $I(\omega)$, and
population $\mathcal{P}_{1}^{\mathrm{ad}}(t)$ due to the neglect of the
residual couplings all remain small (see Fig.~\ref{fig:no3_disp_main_2}); in
particular, quantum fidelity $\mathcal{F}(t)$ remains above $0.95$ for all
times until the final time $t_{f}=50$ n.u. [see panel (e)]. (Note that the exact benchmark wavepackets [in panels~(a) of
Figs.~\ref{fig:no3_disp_main_1} and \ref{fig:no3_disp_main_2}] propagated in
the two different quasidiabatic representations are slightly different
not only because they are displayed in different
representations but also because the initial states are different---they have
the same analytical form but in two different quasidiabatic representations.)

\section{Conclusion}

We have shown that the common practice of neglecting the residual
nonadiabatic couplings between quasidiabatic states can substantially lower
the accuracy of nonadiabatic simulations and that the decrease of accuracy
depends on the system, initial state, and employed quasidiabatization scheme.
One can, therefore, answer the question posed in the title only after a
careful analysis. In Sec.~\ref{sec:numerical_examples}, we have provided
several examples where the approximate quasidiabatic Hamiltonian gives wrong
results. Because it is potentially dangerous to employ an approximation
without evaluating its impact, we have proposed a method to rigorously
quantify the errors caused by ignoring the residual couplings.

When the residual couplings are significant and cannot be
neglected, we suggest performing nonadiabatic simulations with the rarely used
exact quasidiabatic Hamiltonian (\ref{eq:mol_H_accurate}), which not only is
analytically equivalent to the adiabatic Hamiltonian~(\ref{eq:mol_H_full}),
but also yields numerically accurate results regardless of the magnitude of
the residual couplings (as shown in Sec.~S2 of the supplementary material and
in Ref.~\onlinecite{Choi_Vanicek:2020}). Although the general applicability of
the exact quasidiabatic Hamiltonian depends on the availability of residual
nonadiabatic couplings, these can be evaluated by employing recently developed
schemes\cite{Zhu_Yarkony:2012a, Zhu_Yarkony:2014, Zhu_Yarkony:2014a,
Zhu_Yarkony:2016a, Zhu_Yarkony:2016b} even in rather complicated multi-state
systems involving multiple conical intersections (including those between
three electronic states\cite{Coe_Martinez:2005, Schuurman_Yarkony:2006,
Matsika_Yarkony:2002, Matsika:2005, Kistler_Matsika:2008}). In complex systems
where all practical quasidiabatization schemes lead to significant residual
couplings, propagating the wavepacket with the exact quasidiabatic Hamiltonian
would be particularly beneficial. Although the nonseparable form of this
Hamiltonian complicates the time propagation, there exist efficient geometric
integrators, such as the high-order compositions of the implicit midpoint
method used here, which are applicable even to such Hamiltonians.

Last but not least, an accurate propagation of the wavepacket with the exact
quasidiabatic Hamiltonian would be extremely useful for establishing highly
accurate benchmarks in unfamiliar systems, where the impact of the residual
nonadiabatic couplings on the quantum dynamics simulations is not yet known.

\section*{Supplementary material}

See the supplementary material for the details of the numerical
evaluation of the magnitude of the residual couplings (Sec.~S1); demonstration
of the negligibility of spatial and time discretization errors (Sec.~S2);
conservation of geometric properties by the implicit midpoint method
(Sec.~S3); time dependence of position, potential energy, and
distance (Sec.~S4); and importance of the residual couplings for different
Jahn--Teller coupling coefficients and different initial populations (Sec.~S5).

\section*{Acknowledgments}

The authors acknowledge the financial support from the European Research
Council (ERC) under the European Union's Horizon 2020 research and innovation
programme (grant agreement No. 683069 -- MOLEQULE) and thank Tomislav
Begu\v{s}i\'{c} and Nikolay Golubev for useful discussions.

\section*{Data Availability}

The data that support the findings of this study are contained in the paper
and the supplementary material.

\setcounter{section}{0}
\setcounter{equation}{0}
\setcounter{figure}{0}
\setcounter{table}{0}
\renewcommand{\theequation}{S\arabic{equation}}
\renewcommand{\thefigure}{S\arabic{figure}} \renewcommand{\thesection}{S\arabic{section}}

\bigskip

\textbf{\large Supplementary material for: How important are the residual nonadiabatic couplings for an accurate simulation of nonadiabatic quantum dynamics in a quasidiabatic representation?}

\section{\label{sec:R_eval} Numerical evaluation of the magnitude of the
residual couplings}

The evaluation of the magnitude $\mathcal{R}[\mathbf{F}_{\mathrm{qd}}(Q)]$ of
residual couplings requires an integration over the entire nuclear space [see
Eq.~(10) of the main text]. In the main text, we approximate the integral
numerically on a finite grid of $N\times N$ points between $-Q_{l}$ and
$Q_{l}$ for $l\in\{1,2\}$. In the nonadiabatic simulations, we used $N=64$ and
$Q_{l}=10$ n.u. in the NO$_{3}$ model and $N=32$ and $Q_{l}=7$ n.u. in the HCN model.

To evaluate the magnitude $\mathcal{R}[\mathbf{F}_{\mathrm{qd}}(Q)]$ of the
residual nonadiabatic couplings, we have chosen a grid narrower than the one
used for nonadiabatic simulations because $\mathcal{R}[\mathbf{F}%
_{\mathrm{qd}}(Q)]$ evaluated on a wider grid would not be informative, as can
be seen from the following consideration: In addition to the central conical
intersection (at $Q=0$), the cubic Jahn--Teller model has six other conical
intersections at $\rho(Q)=\rho_{+}$ and $\phi(Q)=-2\pi/3,0,2\pi/3$ and at
$\rho(Q)=\rho_{-}$ and $\phi(Q)=-\pi/3,\pi/3,\pi$, where $\rho_{\pm}=(c_{2}%
\pm\sqrt{c_{2}^{2}-4c_{1}c_{3}})/(2c_{3})$. Although the singularities of the
nonadiabatic couplings at these additional conical intersections remain even
after the quasidiabatization by the regularized diabatization scheme, these
singularities are sufficiently far from the region of the dynamics and do not
have a significant effect on the simulations (i.e., numerical convergence was
achieved despite the remaining singularities; see
Sec.~\ref{sec:numerical_errors} of the supplementary material). However,
because it diverges to infinity, the magnitude $\mathcal{R}[\mathbf{F}%
_{\mathrm{qd}}(Q)]$ of the residual couplings evaluated on a grid that
includes these additional conical intersections is not meaningful. We,
therefore, evaluate $\mathcal{R}[\mathbf{F}_{\mathrm{qd}}(Q)]$ on a narrower
grid that does not include these singular residual couplings.

\section{\label{sec:numerical_errors} Negligibility of spatial and time
discretization errors}

For the results presented in the main text to be valid, both the spatial and
time discretization errors should be smaller than the errors due to
the neglect of the residual nonadiabatic couplings. We used distance
functionals $\epsilon_{N}^{(\mathrm{grid})}[\bm{\psi}(t)]:=\Vert
\bm{\psi}^{(\Delta t,N)}(t)-\bm{\psi}^{(\Delta t,2N)}(t)\Vert$ and
$\epsilon_{\Delta t}^{(\mathrm{time})}[\bm{\psi}(t)]:=\Vert\bm{\psi}^{(\Delta
t,N)}(t)-\bm{\psi}^{(\Delta t/2,N)}(t)\Vert$ to measure the spatial and time
discretization errors of $\bm{\psi}^{(\Delta t,N)}(t)$, the molecular
wavepacket propagated to time $t$ with the time step of $\Delta t$ on a grid
of $N\times N$ points. Similarly, we used $\epsilon_{N}^{(\mathrm{grid}%
)}(A):=|A^{(\Delta t,N)}-A^{(\Delta t,2N)}|$ and $\epsilon_{\Delta
t}^{(\mathrm{time})}(A):=|A^{(\Delta t,N)}-A^{(\Delta t/2,N)}|$ to measure the
spatial and time discretization errors of $A^{(\Delta t,N)}$, an observable
$A$ obtained from a simulation on a grid of $N\times N$ points with the time
step $\Delta t$. The grid of $2N\times2N$ points was defined so that it was
both denser and wider by a factor of $\sqrt{2}$ (both in position and momentum
spaces) compared to the grid of $N\times N$ points.

\begin{figure}
[htb!]\includegraphics[]{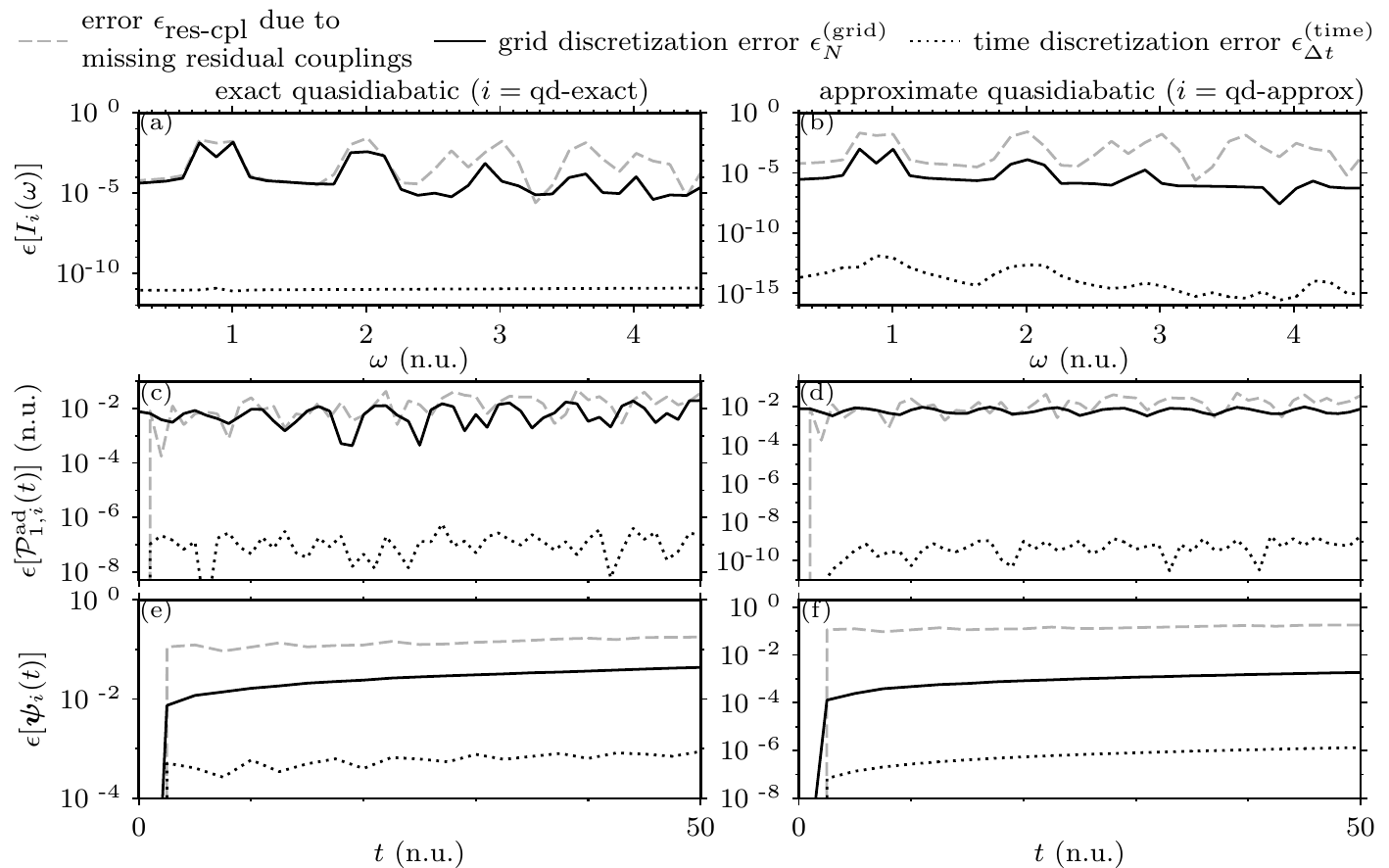}
\caption{Negligibility of spatial and time discretization errors of
the quantities presented in Fig.~2 of the main text:
(a)--(b) power spectrum $I_{i}(\omega)$, (c)--(d) population $\mathcal{P}_{1,i}^{\mathrm{ad}}(t)$, and (e)--(f)
wavepacket $\bm{\psi}_{i}(t)$ obtained with either the exact [$i=\textrm{qd-exact}$, panels (a), (c), (e)] or
approximate [$i=\textrm{qd-approx}$, panels (b), (d), (f)] Hamiltonian. In each panel, errors $\epsilon_{\textrm{res-cpl}}$ due
to the neglect of the residual couplings are compared with the grid discretization errors
$\epsilon_{N}^{\mathrm{(grid)}}$ and time discretization errors $\epsilon_{\Delta t}^{\mathrm{(time)}}$. Note that
$\epsilon_{\textrm{res-cpl}}[\bm{\psi}(t)] \equiv \mathcal{D}(t) :=
\| \bm{\psi}_{\textrm{qd-approx}}(t) - \bm{\psi}_{\textrm{qd-exact}}(t) \| $.}\label{fig:no3_vert_main_error}%

\end{figure}

\begin{figure}
[htb!]\includegraphics[]{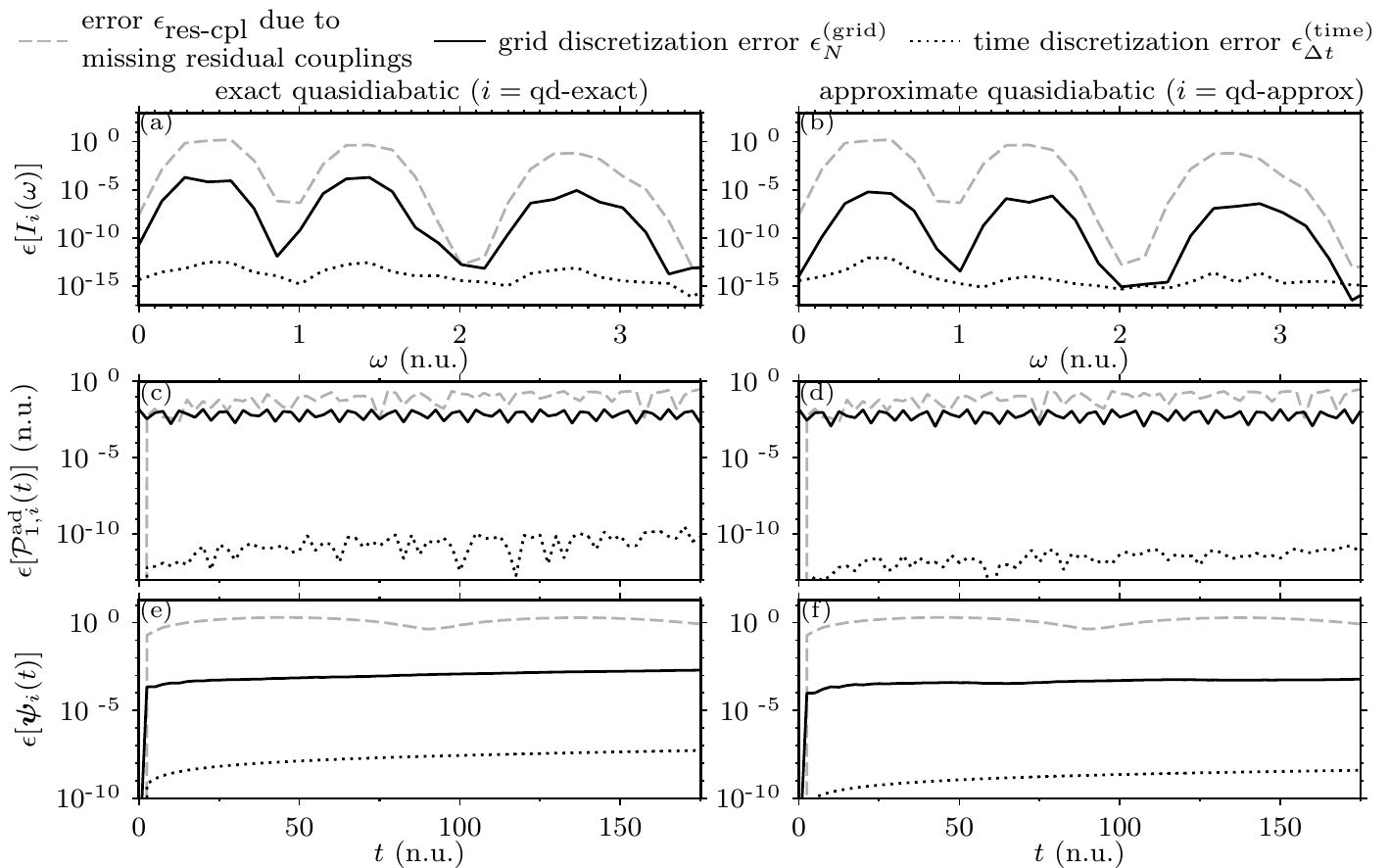}
\caption{Negligibility of spatial and time discretization errors of
the quantities presented in Fig.~4 of the main text. See the caption of Fig.~\ref{fig:no3_vert_main_error} for details.}\label{fig:hcn_main_error}%

\end{figure}

\begin{figure}
[htb!]\includegraphics[]{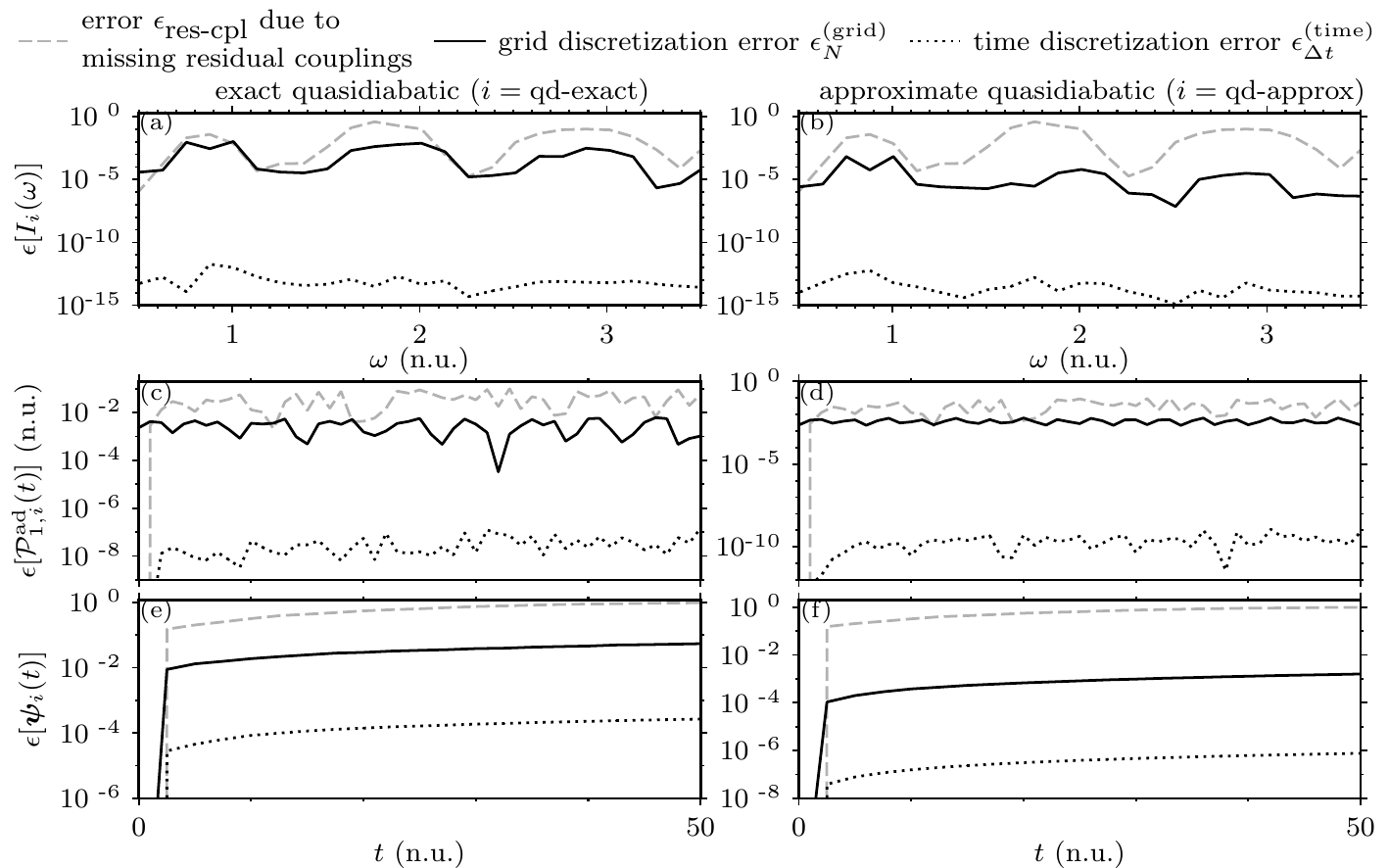}
\caption{Negligibility of spatial and time discretization errors of
the quantities presented in Fig.~5 of the main text. See the caption of Fig.~\ref{fig:no3_vert_main_error} for details.}\label{fig:no3_disp_main_1_error}%

\end{figure}

\begin{figure}
[htb!]\includegraphics[]{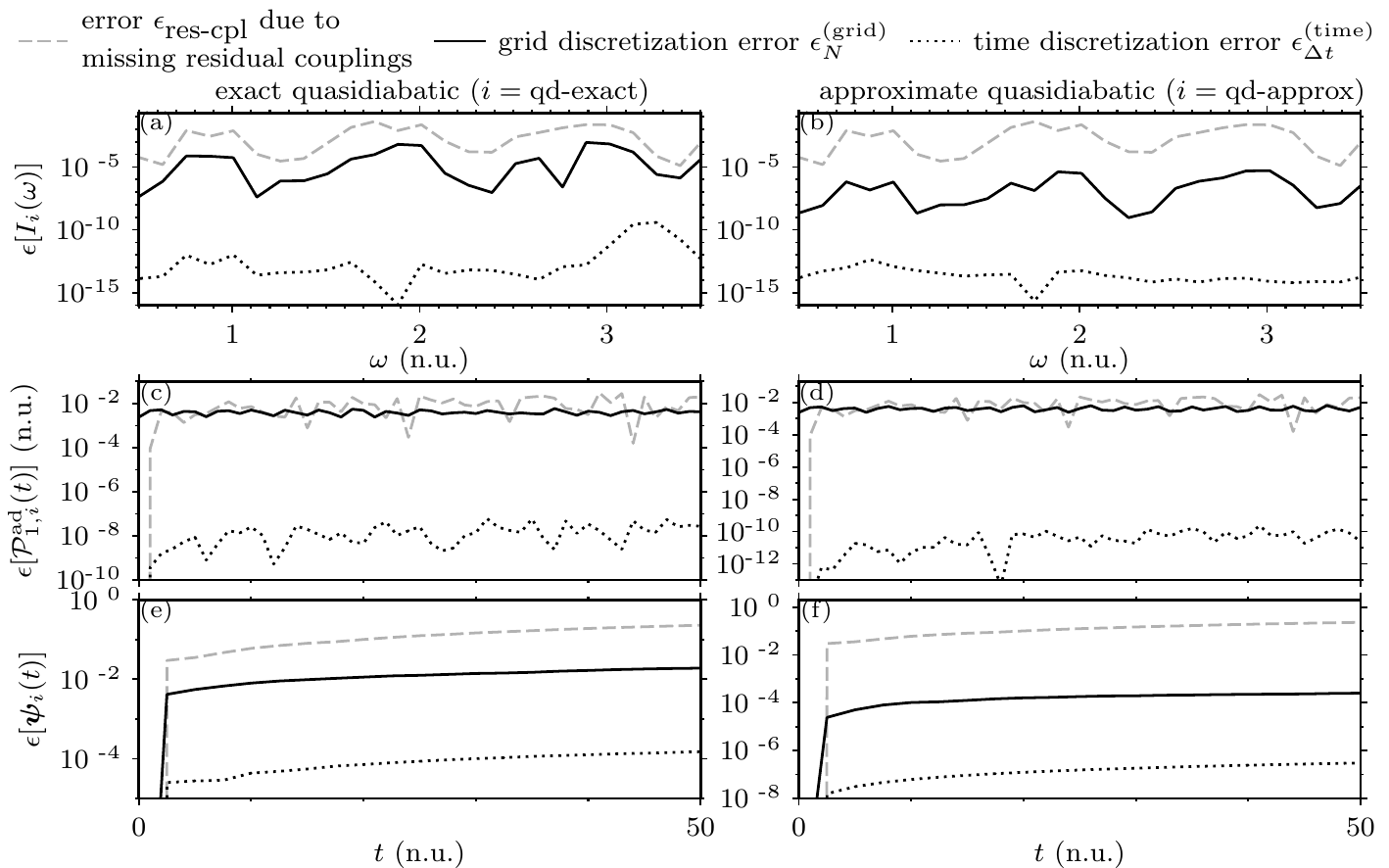}
\caption{Negligibility of spatial and time discretization errors of
the quantities presented in Fig.~6 of the main text. See the caption of Fig.~\ref{fig:no3_vert_main_error} for details.}\label{fig:no3_disp_main_2_error}%

\end{figure}

Figures~\ref{fig:no3_vert_main_error}--\ref{fig:no3_disp_main_2_error} show
that the grid discretization errors of the quantities presented in the main
text are smaller than the errors due to the neglect of the residual
couplings. Moreover, thanks to the high order of accuracy of the employed time
propagation scheme, the time discretization errors are negligible in
comparison with the corresponding spatial discretization errors. The small
numerical errors of the wavepackets propagated with the exact quasidiabatic
Hamiltonian validate them as the reference benchmark wavepackets because the
exact quasidiabatic Hamiltonian is exact in the sense that it is a
coordinate-dependent unitary transform of the adiabatic Hamiltonian.

\section{\label{sec:conserv_geom_prop} Conservation of geometric properties by
the implicit midpoint method}

\begin{figure}
[htb!]\includegraphics[]{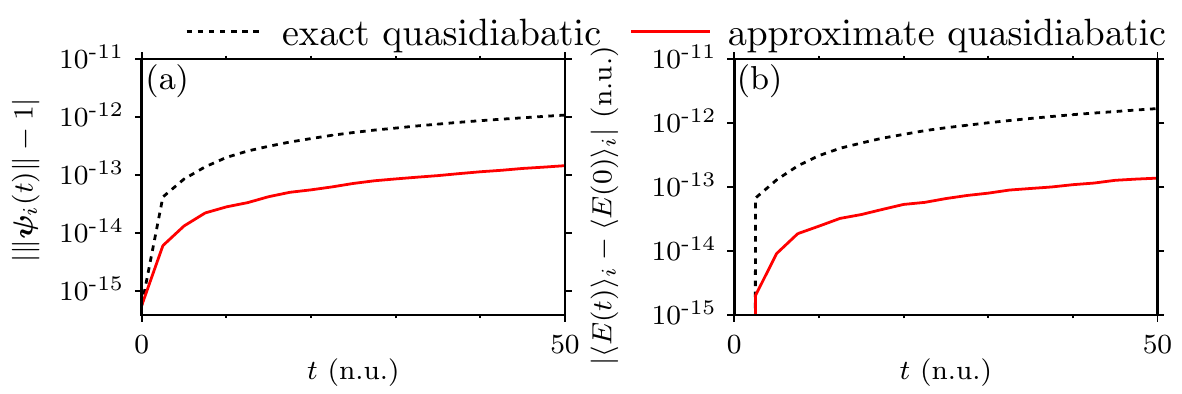}
\caption{Exact conservation of geometric properties in the vertical excitation of NO$_{3}$ (from Sec.~III~A
of the main text) by the employed integrator: the conservation of the (a) norm $\| \bm{\psi}_{i} (t) \|$ and
(b) energy $\langle E(t) \rangle_{i}$ of the wavepacket propagated with either the exact
($i = \textrm{qd-exact}$) or approximate ($i = \textrm{qd-approx}$) quasidiabatic Hamiltonian.
The initial values are $\| \bm{\psi}_{i} (0) \| = 1$ and $\langle E(0) \rangle_{i} = 1$ n.u.}
\label{fig:no3_vert_norm_E}
\end{figure}

Here, we demonstrate the exact conservation of the wavepacket's norm
$\Vert\bm{\psi}(t)\Vert$ and energy $\langle E(t)\rangle$ by the optimal
eighth-order\cite{Kahan_Li:1997} composition\cite{Suzuki:1990, Yoshida:1990,
book_Hairer_Wanner:2006, book_Lubich:2008} of the implicit midpoint method.
Figure~\ref{fig:no3_vert_norm_E} shows that both the norm and energy are
conserved to machine precision ($<10^{-12}$) in the model of
vertical excitation of NO$_{3}$ from Sec.~III~A of the main text. In fact,
they are conserved to machine precision regardless of the size of the time
step (not shown). We refer the reader to Ref.~\onlinecite{Choi_Vanicek:2019}
and the references therein for the analytical proof and numerical
demonstration of the preservation of geometric properties of the exact
solution (the conservation of norm, energy, and inner-product, linearity,
symplecticity, stability, symmetry, and time reversibility) by the
compositions of the implicit midpoint method.

\begin{figure}
[htb!]\includegraphics[]{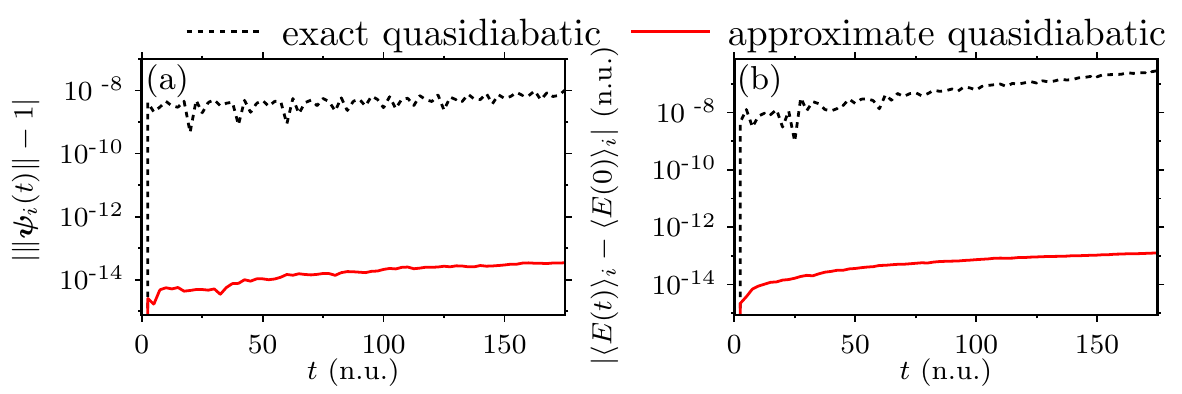}
\caption{Exact conservation of geometric properties in the HCN model (from Sec.~III~B of the main text)
by the employed integrator. See the caption of Fig.~\ref{fig:no3_vert_norm_E} for details. Here, the initial norm
is $\| \bm{\psi}_{i} (0) \| = 1$ and the initial energies are
$\langle E(0) \rangle_{\textrm{qd-exact}} = 0.6$ n.u. and $\langle E(0) \rangle_{\textrm{qd-approx}} = 0.7$ n.u.
} \label{fig:hcn_norm_E}
\end{figure}

Figure~\ref{fig:hcn_norm_E} shows the norm and energy
conservation in the HCN model from Sec.~III~B of the main text. Note that here
the norm is not conserved to machine precision, but \textquotedblleft
only\textquotedblright\ to $10^{-8}$; the subtle reason for this effect is
that on a finite grid, the exact quasidiabatic Hamiltonian~(9) of the main
text is only approximately Hermitian. In contrast, simulations with exactly
Hermitian Hamiltonians [e.g., Hamiltonians~(12) and (28) of the main text]
conserve the norm and energy exactly regardless of the grid density (not
shown).

\section{\label{sec:position_pot_energy_distance} Time dependence of position,
potential energy, and distance}

\begin{figure}
[htb!]\includegraphics[]{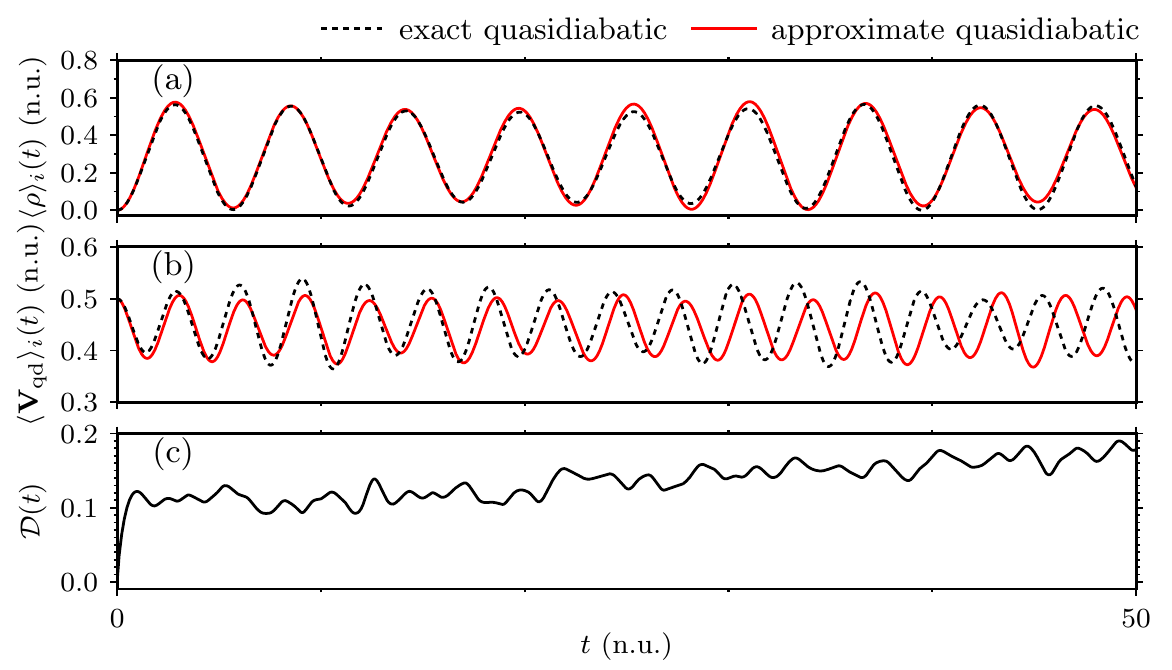}
\caption{Importance of the residual nonadiabatic couplings in the NO$_{3}$ model from
Sec.~III~A of the main text. The figure, which complements Fig.~2 of the main text,
shows the time dependence of (a) position $\langle \rho \rangle_{i}(t)$,
(b) potential energy $\langle \mathbf{V}_{\mathrm{qd}} \rangle_{i}(t)$, and
(c) distance $\mathcal{D}(t)$ [Eq.~(\ref{eq:distancespl})].}\label{fig:no3_vert_spl}%

\end{figure}

\begin{figure}
[htb!]\includegraphics[]{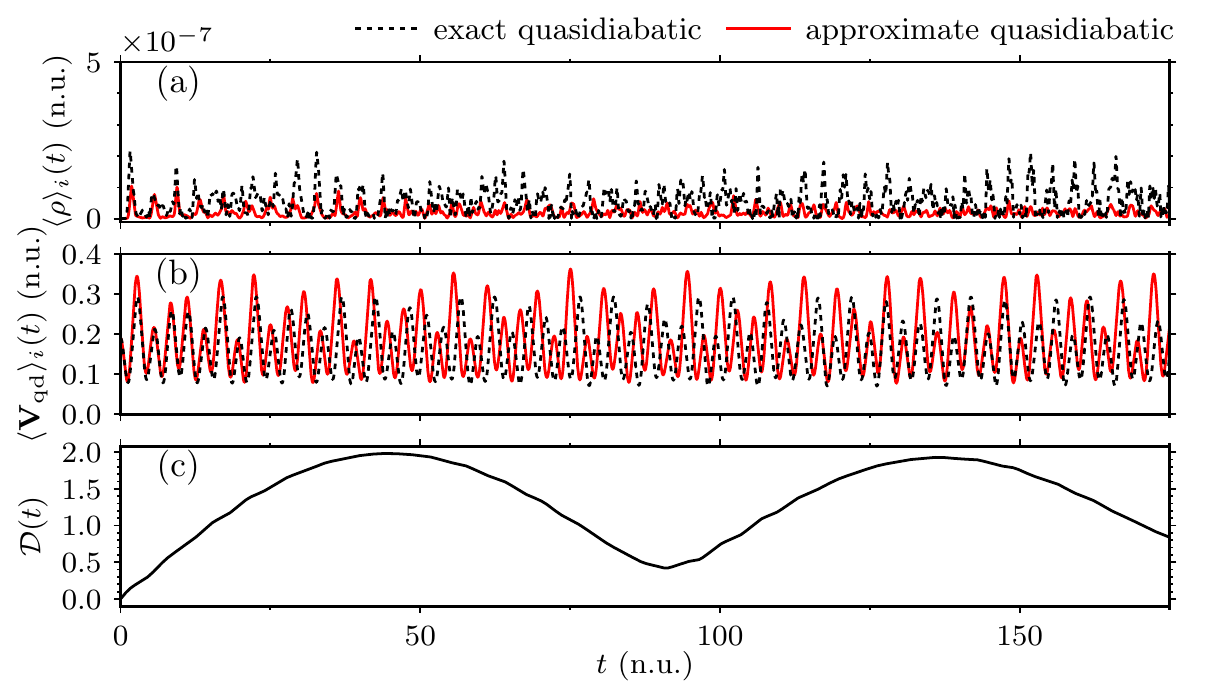}
\caption{Importance of the residual nonadiabatic couplings in the HCN model from
Sec.~III~B of the main text. The figure, which complements Fig.~4 of the main text,
shows the time dependence of (a) position $\langle \rho \rangle_{i}(t)$,
(b) potential energy $\langle \mathbf{V}_{\mathrm{qd}} \rangle_{i}(t)$, and
(c) distance $\mathcal{D}(t)$ [Eq.~(\ref{eq:distancespl})].
Note that the wavepacket remains at $\langle \rho \rangle_{i}(t) = 0$ throughout the dynamics and
only its width changes (not shown).}\label{fig:hcn_spl}
\end{figure}

\begin{figure}
[htb!]\includegraphics[]{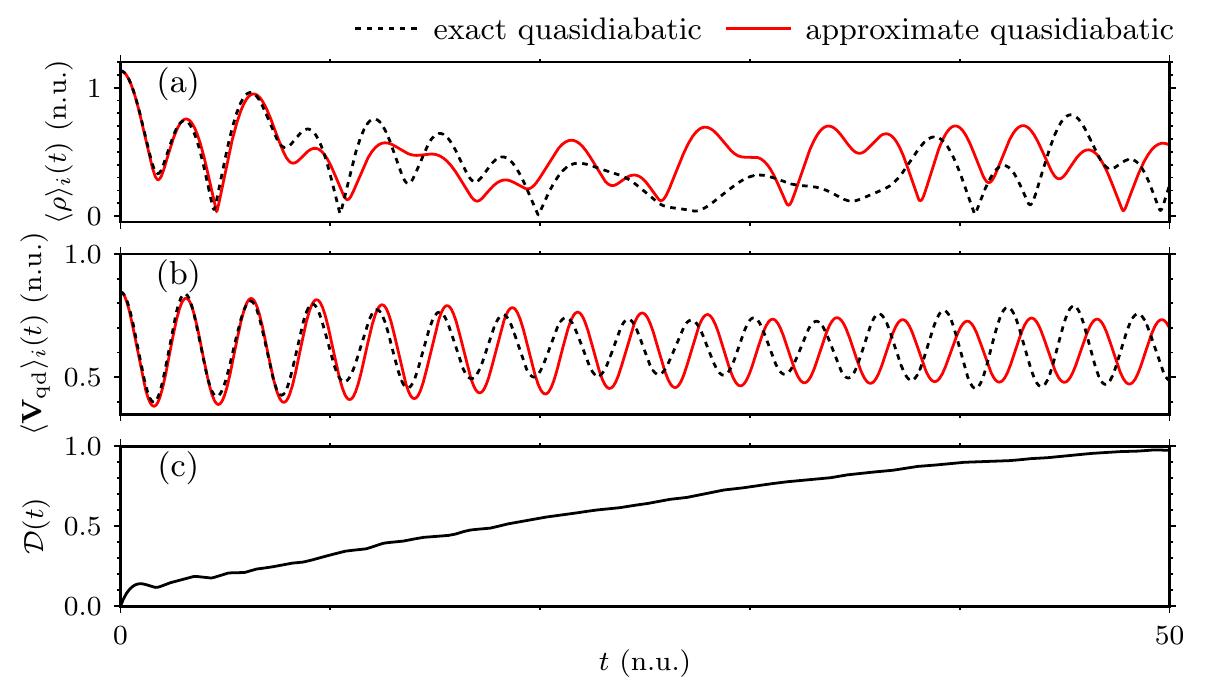}
\caption{Importance of the residual nonadiabatic couplings in the model of a displaced excitation
of NO$_{3}$  from Sec.~III~C of the main text. The molecular Hamiltonian was
quasidiabatized with the first-order ($j=1$) scheme.
The figure, which complements Fig.~5 of the main text, shows the time dependence
of (a) position $\langle \rho \rangle_{i}(t)$,
(b) potential energy $\langle \mathbf{V}_{\mathrm{qd}} \rangle_{i}(t)$, and
(c) distance $\mathcal{D}(t)$ [Eq.~(\ref{eq:distancespl})].}\label{fig:no3_disp_spl_1}%

\end{figure}

\begin{figure}
[htb!]\includegraphics[]{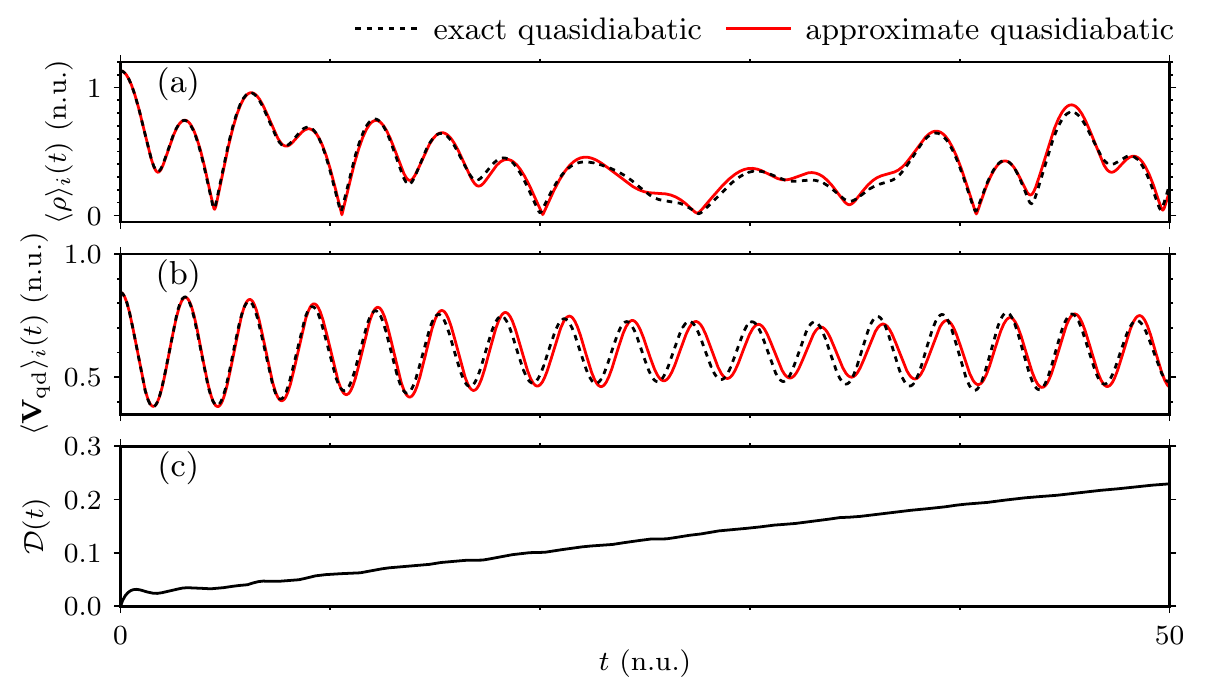}
\caption{Importance of the residual nonadiabatic couplings in the model of a displaced excitation
of NO$_{3}$  from Sec.~III~C of the main text. The molecular Hamiltonian was
quasidiabatized with the second-order ($j=2$) scheme.
The figure, which complements Fig.~6 of the main text, shows the time dependence
of (a) position $\langle \rho \rangle_{i}(t)$,
(b) potential energy $\langle \mathbf{V}_{\mathrm{qd}} \rangle_{i}(t)$, and
(c) distance $\mathcal{D}(t)$ [Eq.~(\ref{eq:distancespl})].}
\label{fig:no3_disp_spl_2}
\end{figure}

To supplement Figs.~2, 4--6 of the main text, we present, in
Figs.~\ref{fig:no3_vert_spl}--\ref{fig:no3_disp_spl_2}, the time dependence of
the position $\langle\rho\rangle_{i}(t):=[\sum_{l=1}^{2}\langle\bm{\psi}_{i}%
(t)|\hat{Q}_{l}|\bm{\psi}_{i}(t)\rangle^{2}]^{1/2}$ [panels~(a)] and potential
energy $\langle\mathbf{V}_{\mathrm{qd}}\rangle_{i}(t):=\langle\bm{\psi}_{i}%
(t)|\mathbf{V}_{\mathrm{qd}}(\hat{Q})|\bm{\psi}_{i}(t)\rangle$ [panels~(b)]
obtained either with the approximate ($i=\text{qd-approx}$) or exact
($i=\text{qd-exact}$) quasidiabatic Hamiltonian. In panels~(c), we show the
distance
\begin{equation}
\mathcal{D}(t):=\Vert\bm{\psi}_{\text{qd-approx}}%
(t)-\bm{\psi}_{\text{qd-exact}}(t)\Vert\label{eq:distancespl}%
\end{equation}
between the wavepackets propagated either with the approximate or exact Hamiltonian.

\section{\label{sec:appendixc}Importance of the residual couplings for
different Jahn--Teller coupling coefficients and different initial
populations}

\begin{figure}
[htb!]\includegraphics[]{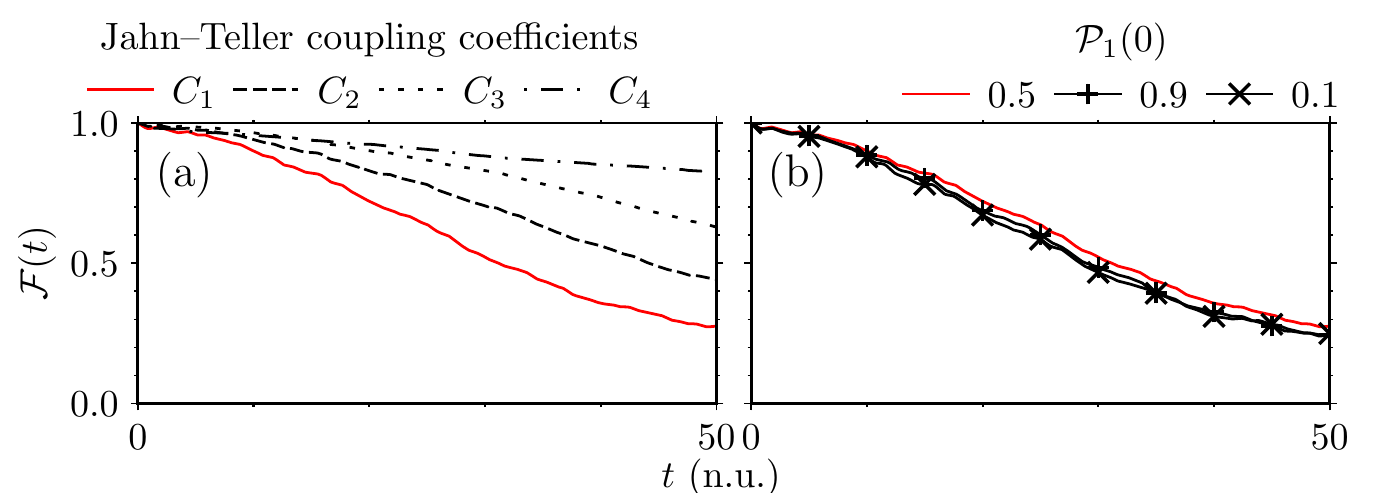} \caption{
Importance of the residual couplings in the displaced excitation of NO$_{3}$ (from Sec.~III~C of the
main text) for (a) different Jahn--Teller coupling coefficients and (b) different initial quasidiabatic populations
$\mathcal{P}_{1}(0)$. The importance of the residual couplings is measured by quantum fidelity [as in Fig.~5(e)
of the main text] and shown only for the first-order quasidiabatization scheme. Results based on the second-order
quasidiabatization scheme are not shown since they are  very accurate [i.e., $\mathcal{F}(t) \approx 1$] in all
presented cases. The red solid lines in the two panels are
identical  to each other and to the black solid line in Fig.~5(e) of the main text.}
\label{fig:param_init}
\end{figure}

In Fig.~\ref{fig:param_init}(a), we consider four sets of Jahn--Teller
coupling coefficients, represented by triples $C=(c_{1},c_{2},c_{3})$ [all in
natural units (n.u.)]: $C_{1}=(0.375,-0.0668,0.0119)$, $C_{2}%
=(0.375,-0.05,-0.0155)$, $C_{3}=(0.375,-0.037,-0.0185)$, and $C_{4}%
=(1.5,0.388,0.052)$. Among these, triple $C_{1}$ consists of the coefficients
of the NO$_{3}$ model discussed in Sec.~III~C of the main text. The other
triples were chosen so that the magnitudes of the residual couplings in the
first-order quasidiabatization were $\mathcal{R}[\mathbf{F}_{\mathrm{qd}%
}(Q)]=3.8$ n.u. for both $C_{1}$ and $C_{4}$, $\mathcal{R}[\mathbf{F}%
_{\mathrm{qd}}(Q)]=2.4$ n.u. for $C_{2}$, and $\mathcal{R}[\mathbf{F}%
_{\mathrm{qd}}(Q)]=1.6$ n.u. for $C_{3}$ and so that $\mathcal{R}%
[\mathbf{F}_{\mathrm{qd}}(Q)]=0.5$ n.u. for all four triples if the
second-order scheme was employed. Because the triples $C_{1}$, $C_{2}$, and
$C_{3}$ are similar, the resulting nonadiabatic dynamics were also similar; in
contrast, the dynamics with triple $C_{4}$ was very different (not shown). On
one hand, panel~(a) of Fig.~\ref{fig:param_init} shows that there is a
positive correlation between the magnitude and importance of the residual
couplings when the dynamics are similar. On the other hand, even when the
magnitudes $\mathcal{R}[\mathbf{F}_{\mathrm{qd}}(Q)]$ are the same, the
importance of residual couplings can differ significantly if the nonadiabatic
dynamics are not the same (compare the results for $C_{1}$ and $C_{4}$).

In panel (b), we consider three different initial states that only differed in
the initial quasidiabatic populations: the population of the first state was
$\mathcal{P}_{1}(0)=0.5$, $0.9$, or $0.1$; in all cases, $\mathcal{P}%
_{2}(0)=1-\mathcal{P}_{1}(0)$. The initial state with $\mathcal{P}_{1}(0)=0.5$
was the one analyzed in Sec.~III~C of the main text. Panel~(b) of
Fig.~\ref{fig:param_init} shows that, in the case of the displaced excitation
of NO$_{3}$, the importance of the residual couplings is almost independent of
the distribution of the initial populations among the different states. This
conclusion, however, may not apply to other systems.

\clearpage

\bibliographystyle{aipnum4-2}
\bibliography{residual_coupling}

\end{document}